%% file: Vortex-dipoles-v5.tex
\documentclass[aps,floatfix,prb,twocolumn,a4paper,10pt,
showpacs,citeautoscript,preprintnumbers,superscriptaddress]{revtex4-1} 

\usepackage[english]{babel}	

\usepackage{amsmath}		
\usepackage{amssymb}		
\usepackage{bbm}			

\usepackage{graphicx}		
\usepackage{graphics}		
\DeclareGraphicsExtensions{.png,.pdf}

\usepackage{color}		
\usepackage[colorlinks=true, pdfstartview=FitV,linkcolor=blue, 
			citecolor=blue, urlcolor=blue]{hyperref}

\makeatletter
\newcommand*{\Equation}{\@ifstar\sEquation\oEquation}
\newcommand{\sEquation}[1]{\begin{equation*}#1\end{equation*}}
\newcommand{\oEquation}[2]{  \begin{equation}\label{#1}#2\end{equation} }
\makeatother

\newcommand{\Align}[2]{\begin{align}\label{#1}#2\end{align}}
\newcommand{\SubAlign}[2]{\begin{subequations}\label{#1}\begin{align}#2\end{align}\end{subequations}}

\newcommand{\bs}{\boldsymbol}

\newcommand{\Figref}[1]{Fig.~\ref{#1}}
\newcommand{\Eqref}[1]{\eqref{#1}}

\newcommand{\eg}{{\it e.g.~}}
\newcommand{\ie}{{\it i.e.~}} 
\newcommand{\resp}{{\it resp.~}} 

\newcommand{\Uone}{\mbox{U(1)}}

\newcommand{\Relative}{\mathbbm{Z}}
\newcommand{\Real}{\mathbbm{R}}

\newcommand{\dd}{\text{d}}
\newcommand{\Exp}[1]{\text{e}^{#1}}

\renewcommand\Im{\mathrm{Im}}

\newcommand{\Q}{\mathcal{Q}}
\newcommand{\F}{\mathcal{F}}
\newcommand{\G}{\mathcal{G}}

\newcommand{\Grad}{{\bs \nabla}}
\newcommand{\Div}{{\bs \nabla}\cdot}
\newcommand{\Curl}{{\bs \nabla}\times}

\newcommand{\oo}{{(1)}}
\newcommand{\ot}{{(2)}}

\newcommand{\x}{{\bs x}}
\newcommand{\ez}{{\bs e}_z}
\newcommand{\et}{{\bs e}_\theta}
\newcommand{\D}{{\bs D}}

\newcommand{\A}{{\bs A}}
\newcommand{\B}{{\bs B}}
\newcommand{\J}{{\bs J}}

\newcommand{\psio}{\psi_{1}}
\newcommand{\psioc}{\psi^{*}_{1}}
\newcommand{\psit}{\psi_{2}}
\newcommand{\psitc}{\psi^{*}_{2}}
\newcommand{\psia}{\psi_{a}}
\newcommand{\psiac}{\psi^{*}_{a}}
\newcommand{\psib}{\psi_{b}}
\newcommand{\psibc}{\psi^{*}_{b}}

\newcommand{\varphiot}{\varphi_{12}}
\newcommand{\nuot}{\nu}
\newcommand{\gammaot}{\gamma}

\begin{document}

\title{Skyrmions induced by dissipationless drag in 
\texorpdfstring{$\Uone\times\Uone$}{U(1)xU(1)} superconductors}

\author{Julien~Garaud} 
\affiliation{Department of Physics, University of Massachusetts Amherst, MA 01003 USA}
\affiliation{Department of Theoretical Physics, Royal Institute of Technology, Stockholm, SE-10691 Sweden}
\author{Karl~A.~H.~Sellin}
\affiliation{Department of Theoretical Physics, Royal Institute of Technology, Stockholm, SE-10691 Sweden}
\author{Juha~J\"aykk\"a} 
\affiliation{Nordita, KTH Royal Institute of Technology and Stockholm University, Stockholm, SE-10691 Sweden}
\author{Egor~Babaev} 
\affiliation{Department of Theoretical Physics, Royal Institute of Technology, Stockholm, SE-10691 Sweden}
\affiliation{Department of Physics, University of Massachusetts Amherst, MA 01003 USA}
\date{\today}

\begin{abstract}
Rather generically, multicomponent superconductors and superfluids 
have intercomponent current-current interaction. 
We show that in superconductors with substantially 
strong intercomponent drag interaction, the topological defects 
which form in external field are characterized by a skyrmionic 
topological charge. We then demonstrate that they can be distinguished 
from ordinary vortex matter by a very characteristic magnetization 
process due to the dipolar nature of inter-skyrmion forces.
The results provide an experimental signature to confirm or rule out
the formation $p$-wave state with reduced spin stiffness in 
$p$-wave superconductors.

\end{abstract}


\maketitle

In multicomponent superconductors and superfluids, the intercomponent 
current-current interaction is rather generic. It usually assumes the 
form of the scalar product of supercurrents in the two components 
$\F_d \propto \J_1 \cdot \J_2 $. 
This kind of interaction between components can have various microscopic 
origins. It was discussed in connection with ${}^3$He-${}^4$He mixtures 
\cite{Andreev.Bashkin:75}; components of order 
parameters of spin-triplet superfluids and superconductors 
\cite{Leggett:68,leggett75,chung1,chung2}; hadronic 
superfluids in neutron stars \cite{Sjoberg:76,Chamel:08,Alpar.Langer.ea:84,
Alford.Good:08,knotsns}; in metallic hydrogen and deuterium \cite{prb09,
Herland.Babaev.ea:10}; in ultracold atomic mixtures \cite{fil,bruder} and 
strongly correlated atomic mixtures in optical lattices 
\cite{Kuklov.Svistunov:03}. In the later case it was shown that it could 
be tuned to have arbitrary strength (in relative units) 
\cite{Kuklov.Svistunov:03}. This kind of interaction for example affects 
rotational response of neutron stars \cite{Alpar.Langer.ea:84} and phase 
transitions, phase diagrams and rotational response of superfluid mixtures 
\cite{Kuklov.Prokofev.ea:04,Kuklov.Prokofev.ea:04a,Dahl.Babaev.ea:08b,
Dahl.Babaev.ea:08a,Dahl.Babaev.ea:08,annals,Herland.Babaev.ea:10}.
Despite the generic character of such interaction, much less is known 
about its effect on the properties of topological excitations and magnetic 
response, beyond the simplest London approximation. Especially, little is 
known about collective properties of such defects. Here, we address this 
problem. We show that beyond a certain interaction threshold, the topological 
defects in the system acquire a skyrmionic topological charge. 
This results in long-range inter-skyrmion forces which alter dramatically
the collective behaviour of vortex matter. 

Note that current-current interaction is fourth order in the order parameters 
densities and second order in their derivatives. Importantly, it is not 
positively defined. Because the total free energy is positively defined, 
the drag term should come with other high-powers terms consistent with the 
$\Uone\times\Uone$ symmetry. Details of the model and how it relates to 
usual London models are discussed in Appendix~\ref{AppTheory}. The precise 
form of these terms is not principally important for the purpose of this 
work, so we investigate a minimal Ginzburg-Landau model (GL), which is 
positively defined and has the correct London limit \cite{Andreev.Bashkin:75} 
\SubAlign{FreeEnergy}{
 \F&= 
\frac{\B^2}{2}  
   +\sum_{a=1,2}\frac{1}{2}|\D\psi_a|^2   
   + \alpha_a|\psi_a|^2+\frac{1}{2}\beta_a|\psi_a|^4 	\label{GL} \\
&+\frac{\nuot}{2}\left|	\Im(\psioc\D\psio)+\Im(\psitc\D\psit)\right|^2	
						      \label{Andreev-Bashkin}\,.
}
Here, $\psi_a=|\psi_a| e^{i\varphi_a}$ are complex fields representing 
the independently conserved superconducting condensates denoted by indices 
$a=1,2$. The term \Eqref{Andreev-Bashkin} contains the intercomponent 
current interaction, as well as higher order terms which makes the free 
energy bounded from below. Besides the drag interaction, the condensates 
are  coupled by electromagnetic interactions in the kinetic terms 
$\D=\nabla+ie\A$. We set the Cooper pair charge as twice the electronic 
charge, then  in these units the coupling constant $e$ parametrizes the 
London penetration length of the magnetic field $\B=\bs\nabla\times\A$, 
and the supercurrent reads as
$
\J\equiv\sum_{a}\J_a= 
   \left[1+\nuot\sum_b|\psi_b|^2\right]\sum_{a}\Im(\psi_a^*\D\psi_a).
$
In connection with spin-triplet systems such models are discussed 
in the situations where the variations of the relative phase 
$\varphi_2-\varphi_1$ of the condensates variations are associated 
with spin degrees of freedom. The drag interaction is then associated
with the  spin stiffness \cite{Leggett:68,leggett75}.  
\begin{figure*}[!htb]
 \hbox to \linewidth{ \hss
 \includegraphics[width=.69\linewidth]{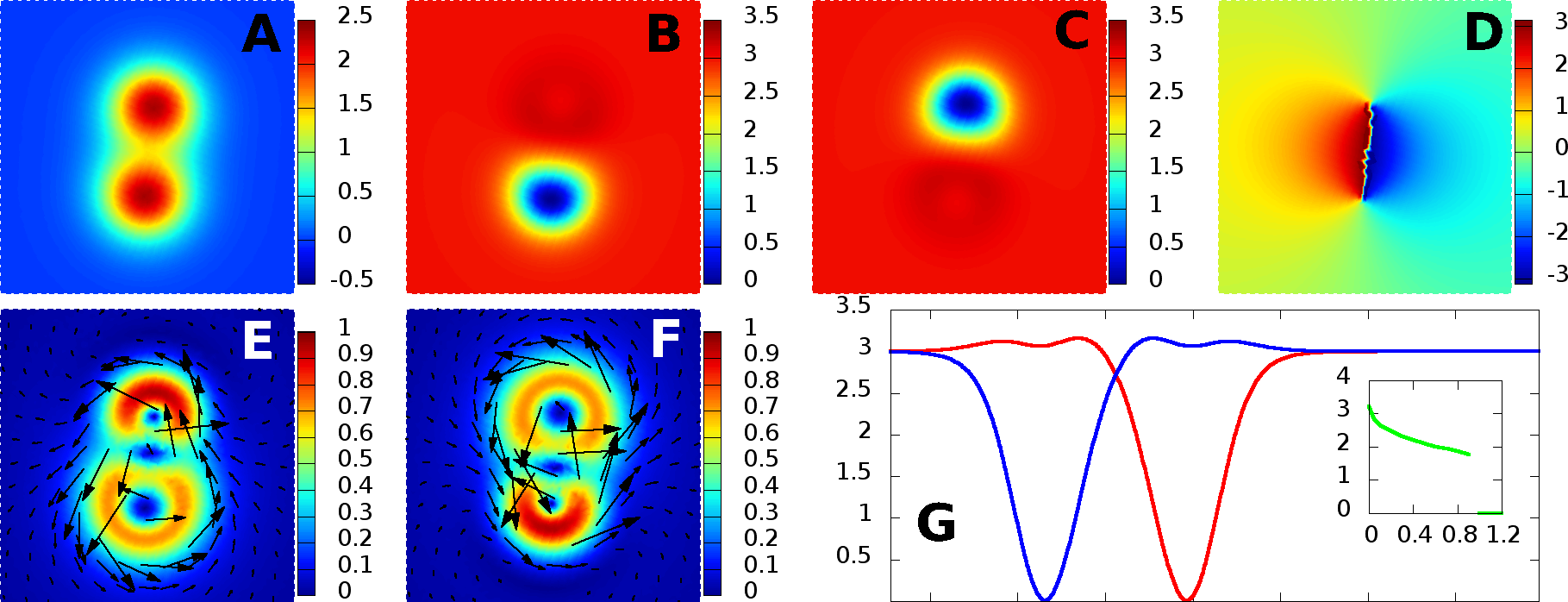}
\hspace{0.1cm}
 \includegraphics[width=.265\linewidth]{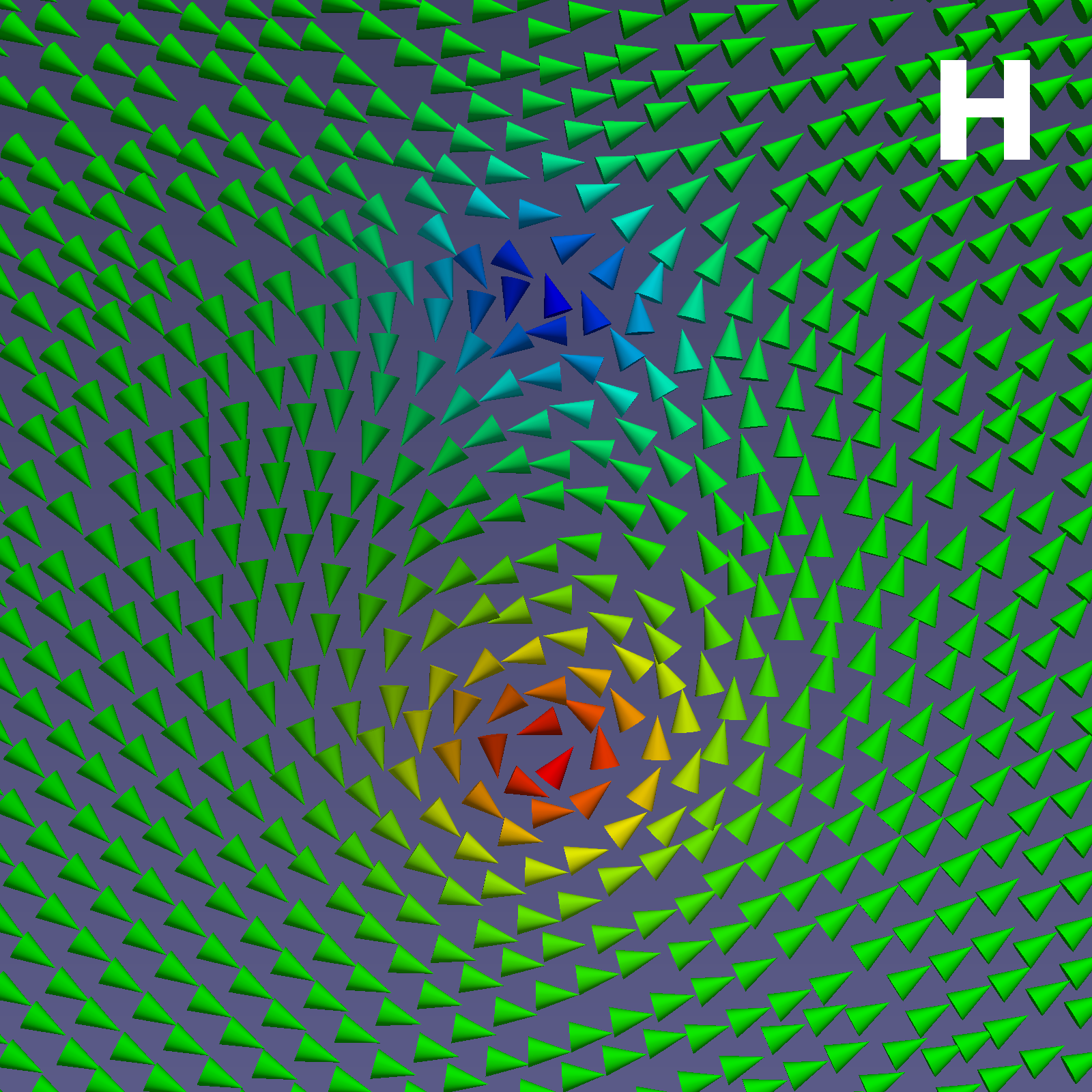}
 \hss}
\caption{
(Color online) A bound state of fractional vortices with $e=0.2$ and 
the potential parameters $(\alpha_a,\beta_a)=(-3.0,1.0)$. The drag 
coupling is $\nuot=2.0$. 
Displayed quantities are the magnetic flux (A) and the densities of 
superconducting condensate $|\psi_1|^2$ (B) and $|\psi_2|^2$ (C). 
(D) shows the phase difference $\varphi_2-\varphi_1$, while individual 
currents $|\J_1|$ and $|\J_2|$ are respectively displayed in (E) and (F). 
Both currents circulate around each core due to drag effect.
(G) shows cross section of densities $|\psi_1|^2$ (red) $|\psi_2|^2$ (blue)
along the $y$ axis. Note the deformed \emph{w-shaped} modulation of 
densities above singularities of the other condensate.
The inset shows the distance between cores as a function of the 
Josephson coupling. At sufficiently strong such coupling skyrmions 
collapse.
The rightmost panel (H) displays the normalized projection of the 
psuedo-spin ${\bf n}$ onto the plane, while colors give the magnitude 
of $n_z$. Blue corresponds to the south pole ($-1$) while red is the north 
pole ($+1$) of the target sphere $S^2$. 
}
\label{Fig:Dipole1}
\end{figure*}

In this work we consider a two-dimensional model. The discussions thus also 
apply to three-dimensional systems invariant along the direction normal to 
the plane. 
The elementary topological excitations of the model are fractional vortices. 
These are field configurations with a $2\pi$ phase winding only in one phase 
(\eg $\varphi_1$ has $\oint \bs\nabla\varphi_1=2\pi$ winding while 
$\oint \bs\nabla\varphi_2=0$). A fractional vortex in the $a$ condensate, 
carries a fraction of flux quantum 
$
\Phi_a	=\oint \A \dd\bs\ell 
		= \frac{|\psi_a|^2}{\varrho^2}\frac{1}{e} \oint\Grad\varphi_a 
      =  \frac{|\psi_a|^2}{\varrho^2}\Phi_0
$ 
with the flux quantum $\Phi_0=2\pi/e$ and the total superfluid density 
$\varrho^2=\sum_a|\psi_a|^2$. Note that this flux quantization is the 
same as in two-component superconductors without drag \cite{frac}. 
Fractional vortices have logarithmically divergent energy. However, 
a composite vortex being the bound state of fractional vortices in both 
condensates (each phase $\varphi_a$ winds $2\pi$) has finite energy and 
carries an integer flux \cite{frac} (see details of the derivation in 
Appendix~\ref{AppTheory}).
In the London limit of a $\Uone\times\Uone$ superconductor, 
fractional vortices can be described by point-like particles 
interacting through logarithmic two-dimensional Coulomb and 
Yukawa interactions, which reads in the general case 
(see details of the derivation in Appendix~\ref{AppTheory})
\Align{Interaction}{
E_{11}&=\ln\frac{R}{x}+wmK_0\left(\frac{x}{\lambda}\right) ,
E_{22}=\ln\frac{R}{x}+\frac{w}{m}K_0\left(\frac{x}{\lambda}\right),\nonumber \\
E_{12}&=-\ln\frac{R}{x}+wK_0\left(\frac{x}{\lambda}\right) \,.
}
Here the interacting energies $E_{ab}$, between vortices in the 
$a$ and $b$ condensates, are expressed in units of 
$2\pi|\psi_1|^2|\psi_2|^2/\varrho^2$. $K_0$ is the modified Bessel 
of second kind and $R$ denotes the system size while the parameters 
$m$ and $w$ are $m =\frac{|\psi_1|^2}{|\psi_2|^2}$ and 
$w =1+\nuot\varrho^2$. $\lambda=\frac{1}{e\sqrt{w\varrho^2}}$ is the 
penetration length of the magnetic field.
For vanishing drag ($w=1$) the  minimum energy corresponds to an 
axially symmetric state of two co-centred fractional vortices 
\cite{frac}. There the Coulomb and Yukawa contributions in $E_{12}$ 
interaction compensate at $x=0$ \cite{npb}.
The drag term \Eqref{Andreev-Bashkin}  (\ie when $w>1$) penalizes 
co-directed currents so the Coulomb and Yukawa contributions of the 
interacting energy $E_{12}$ no longer cancel at  $x=0$ but at some 
finite separation.
In the case of half-quantum vortices this process was studied in 
detail in London model \cite{chung1}.

Here we investigate the structure of single- and multi-vortex states, 
beyond the London limit.  To this end we numerically minimize the free 
energy \Eqref{FreeEnergy} within a finite element framework provided by 
the {\tt Freefem++} library \cite{Hecht:12}. See technical 
details in Appendix~\ref{AppNumerics}).
We find that in contrast to the London limit, weak drag does not produce 
numerically detectable splitting of vortex cores. This is connected with 
the existence of finite cores where the current is modulated by a density 
suppression. Larger drag splits a composite vortex into a bound state of 
well separated fractional vortices.
This is shown on \Figref{Fig:Dipole1}. Note that a single fractional 
vortex has non trivial structure. In particular its magnetic field is 
not exponentially localized and can exhibit flux inversion \cite{juha09}. 
\Figref{Fig:Dipole1} shows that some of the features of isolated fractional 
vortices, reported in\cite{juha09} such as w-shaped modulation of densities, 
are preserved in the split composite vortex.

In general in mulicomponent superconductors there could be terms which 
break the $\Uone\times\Uone$ symmetry explicitly. A typical example is 
$-\eta|\psi_1||\psi_2|\cos\varphi_{12}$. Such terms result in 
asymptotically linear confinement of fractional vortices. We find that 
when such terms are not very strong, the splitting of cores is still 
present as shown on (see \Figref{Fig:Dipole1}-G).
In such a case  dipolar forces are still present, but suppressed at 
the Josephson length.

The bound state of well separated fractional vortices is a \emph{skyrmion}. 
This follows from mapping the two-component model \Eqref{FreeEnergy} 
to an easy-plane non-linear $\sigma$-model \cite{bfn,prb09}. There, 
the pseudo-spin unit vector $\bf n$ is the projection of 
superconducting condensates on spin-$1/2$ Pauli matrices $\bs\sigma$: 
${\bf n}=\frac{\Psi^\dagger\bs \sigma\Psi}{\Psi^\dagger\Psi}$
where $\Psi^\dagger=(\psi_1^*,\psi_2^*)$. When there is non-zero 
drag, the free energy \Eqref{FreeEnergy} can  be written in $\bf n$ 
representation as
\Align{NLSM}{
 \F&= \frac{1}{2}(\Grad\varrho)^2
	+\frac{\varrho^2}{8}\partial_i n_a\partial_i n_a
  	+\frac{\J^2}{2e^2w\varrho^2}+V(\varrho,n_z)	\nonumber \\
	+&\frac{1}{2e^2}\left[\varepsilon_{ijk}\left(
	\partial_i \left(\frac{J_j}{ew\varrho^2}\right)
	-\frac{1}{4}\varepsilon_{abc}n_a\partial_i n_b\partial_j n_c 
	\right)\right]^2
 \,,
}
\begin{figure}[!htb]
 \hbox to \linewidth{\hss 
 \includegraphics[width=\linewidth]{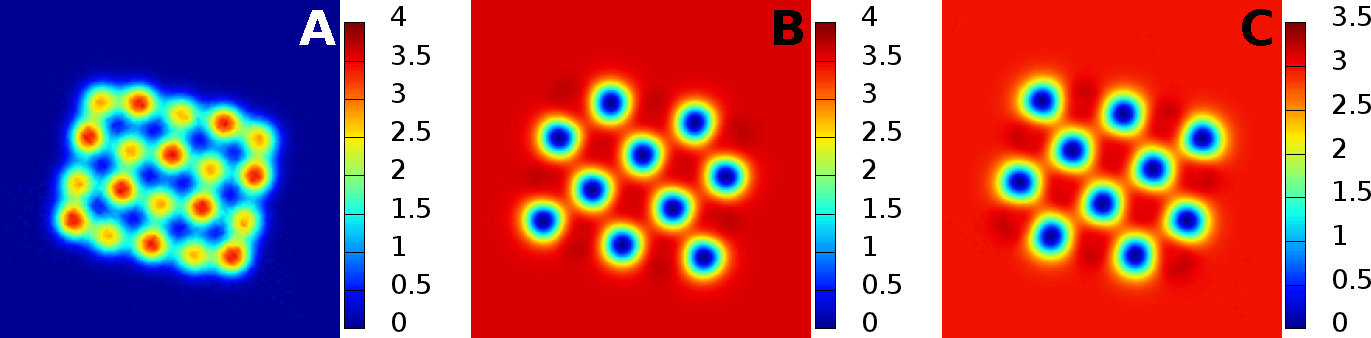}
 \hss} \vspace{0.1cm} 
 \hbox to \linewidth{ \hss
 \includegraphics[width=\linewidth]{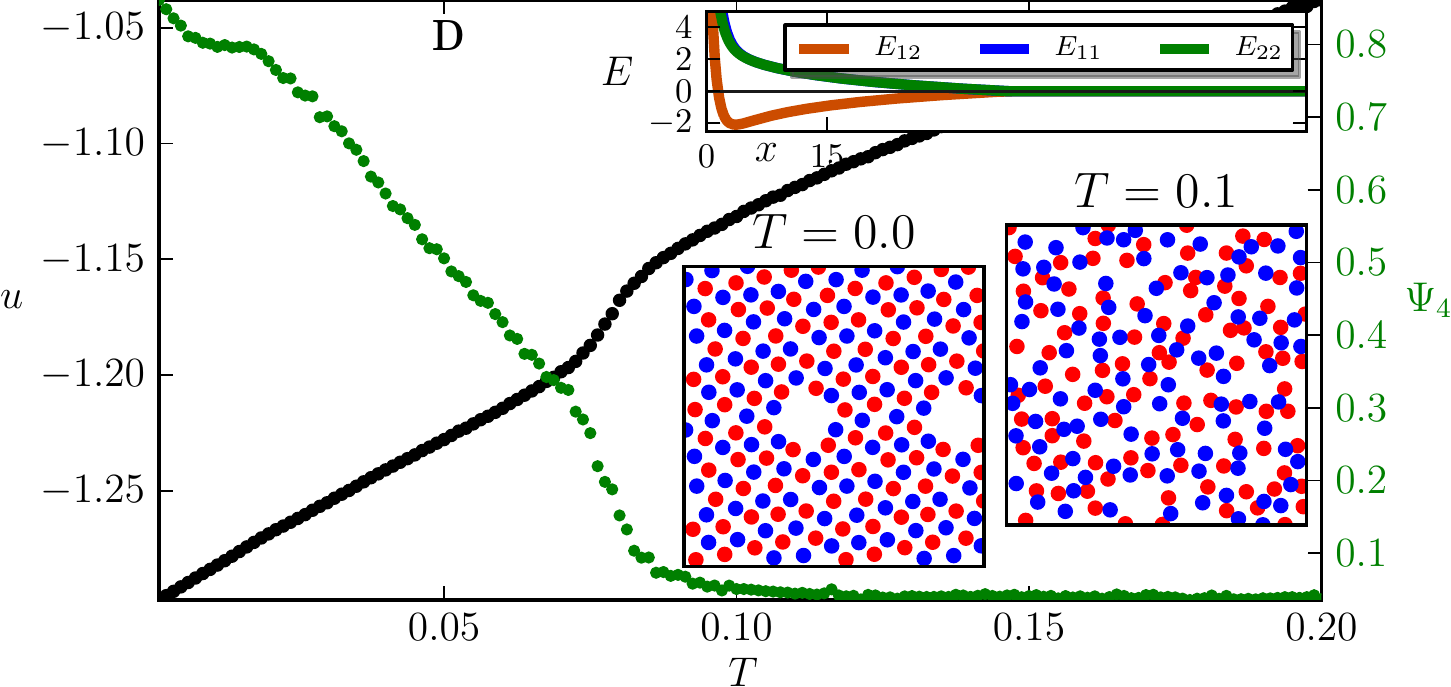}
 \hss}
\caption{
(Color online) A $\Q=10$ quanta configuration bound by dipolar forces. 
Parameters are $(\alpha_1,\beta_1)=(-3.6,1.0)$, 
$(\alpha_2,\beta_2)=(-3.0,1.0)$ with $e=0.6$ and the drag coupling 
$\nuot=2.0$. 
Displayed quantities are the magnetic flux (A) and the densities 
of superconducting condensate $|\psi_1|^2$ (B) and $|\psi_2|^2$ (C).
Lower panel (D) shows a Monte Carlo simulation of hundred point 
particles of each kind interacting according to \Eqref{Interaction}, 
with $0.036$ particles per surface area which allows to emulate the 
skyrmionic lattice melting process. Here, $u$ is the total interaction 
energy per particle and $\Psi_4$ is the square lattice order parameter 
(see Appendix~\ref{AppNumerics}). They show a continuous melting transition 
to a state which has no square lattice ordering but still has bound pairs. 
Insets show low and high temperature states, as well as the interaction 
energies. 
}
\label{Fig:Compact}
\end{figure}
where $\varepsilon$ is the Levi-Civita symbol and $V$ stands for the 
potential terms in \Eqref{GL} (see Appendix~\ref{AppTheory}, for details 
of this derivation).  The pseudo-spin is a map ${\bf n}: S^2\to S^2$, 
classified by the homotopy class $\pi_2(S^2)\in\Relative$, thus defining 
the integer valued topological (skyrmionic) charge  
$
   \Q({\bf n})=\frac{1}{4\pi} \int_{\Real^2}
   {\bf n}\cdot\partial_x {\bf n}\times \partial_y {\bf n}\,\,
  \dd x \dd y 
$. 
Ordinary (composite) vortices with a single core $\Psi=0$, have
$\Q=0$. Here the core-split vortices have non-trivial skyrmionic 
charge $\Q=N$, the number of flux quanta. The quantization of $\Q$ 
follows from the flux quantization, and $\Phi=\Q\Phi_0$ as long as 
cores are split ($\Psi\neq0$).

The calculated pseudo-spin texture of $\bf n$ is shown on panel (H) in 
\Figref{Fig:Dipole1}. Numerically calculated topological charge was found 
to be integer (with a negligible error of order $10^{-4}$)
\footnote{
Note that the topological charge is integer only for when a skyrmion is 
sufficiently far from boundaries. Since when simulating a finite sample in 
applied field, there are states where only part of the skyrmion texture 
enters the sample, in general the topological charge $\Q$ will not be integer.
}.
Note that these skyrmions are quite different from the skyrmions or 
non-axially symmetric vortices considered in superconducting states 
with different number of components and symmetries \cite{PhysRevLett.82.1261,
PhysRevLett.98.187002,knigavko2,PhysRevB.79.014517,
Garaud.Babaev:12,Garaud.Carlstrom.ea:13,*Garaud.Carlstrom.ea:11,
Kobayashi.Nitta:13,sauls}. 
In particular the structural differences in these skyrmions dictate 
different inter-skyrmion forces. This warrants investigation of a state 
of such a superconductor in external field, which we address in the 
following.

\begin{figure*}[!htb]
 \hbox to \linewidth{ \hss
\includegraphics[width=.150\linewidth,height=.158\linewidth]{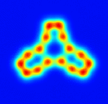}
\includegraphics[width=.150\linewidth,height=.158\linewidth]{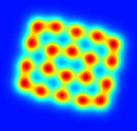}
\includegraphics[width=.150\linewidth,height=.158\linewidth]{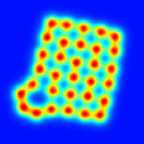}
\includegraphics[width=.150\linewidth,height=.158\linewidth]{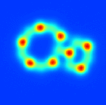}
\includegraphics[width=.150\linewidth,height=.158\linewidth]{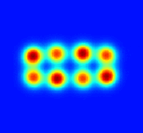}
 \hss}
\caption{(Color online)-- 
Profile of the magnetic field for various bound states of vortices in 
the model \Eqref{FreeEnergy}, carrying $N=8,10,16,8$ and $4$ flux 
quanta respectively. The corresponding potential parameters and 
details of the other physical quantities are given in 
Appendix~\ref{AppAdditional}.
Note that some regimes have extra bi-quadratic density potential term
($\sim|\psi_1|^2|\psi_2|^2$) which is not essential but enriches the 
observed structures. 
}
\label{Fig:Selection}
\end{figure*}
The mapping of fractional vortices to Coulomb charges \Eqref{Interaction} 
suggests that there will be asymptotically power-law inter-Skyrmion 
dipolar interaction forces (attractive for certain orientations and 
repulsive for other). Indeed the long-range Coulomb interaction originates 
in the phase difference mode $\varphiot\equiv\varphi_2-\varphi_1$ 
\cite{npb}. For the pair of fractional vortices it has a clear dipole-like 
structure shown on \Figref{Fig:Dipole1}-(D). The total interaction forces, 
beyond the London limit do not reduce to Coulomb and Yukawa forces and 
are especially complicated at shorter distances due to the presence of 
density modes and Skyrme terms in \Eqref{NLSM}.
\begin{figure}[!hb]
 \hbox to \linewidth{ \hss
 \includegraphics[width=\linewidth]{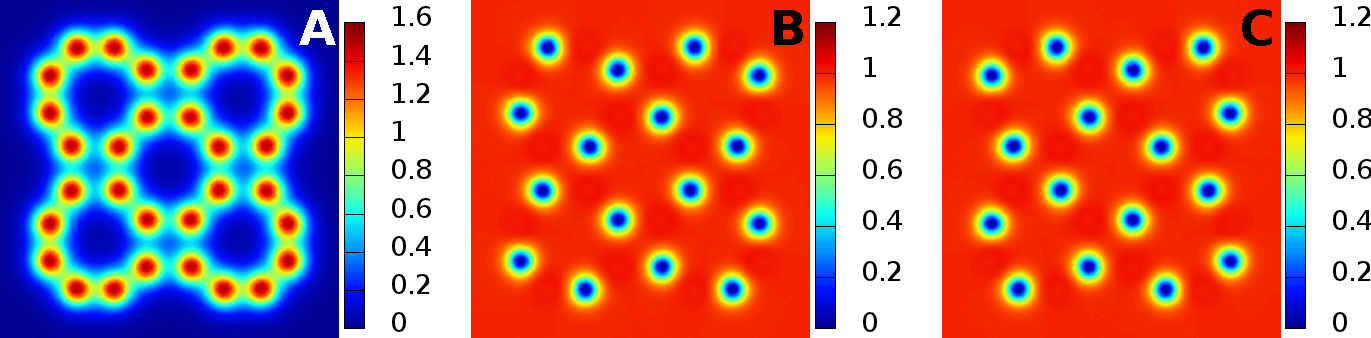}
 \hss}
   \vspace{0.1cm}
 \hbox to \linewidth{ \hss
 \includegraphics[width=\linewidth]{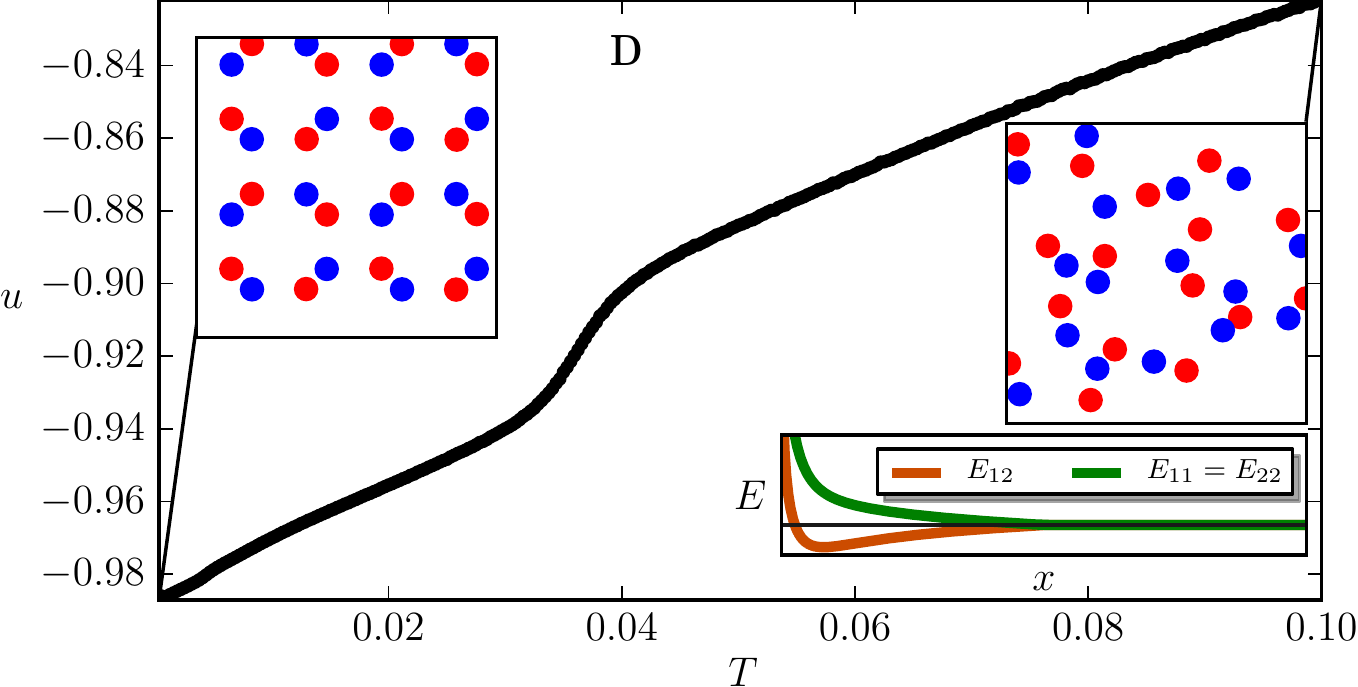}
 \hss}
\caption{
(Color online) A structure carrying $\Q=16$ flux quanta. The 
elementary cell here is a $\Q=4$ skyrmion. The parameters are 
$(\alpha_a,\beta_a)=(-5.0,5.0)$ with $e=0.6$ and the drag coupling 
$\nuot=2.0$. Displayed quantities are the same as in \Figref{Fig:Compact}.
Lower panel shows (D) shows a Monte Carlo simulation of sixteen particles 
of each kind for $0.036$ particles per surface area. In the low temperature 
phase the fractional vortices are paired and ordered in a lattice, and for 
higher temperature the lattice melts but the vortices are still paired. 
}
\label{Fig:Honeycomb}
\end{figure}
\begin{figure*}[!htb]
 \hbox to \linewidth{ \hss
\includegraphics[width=.19\linewidth]{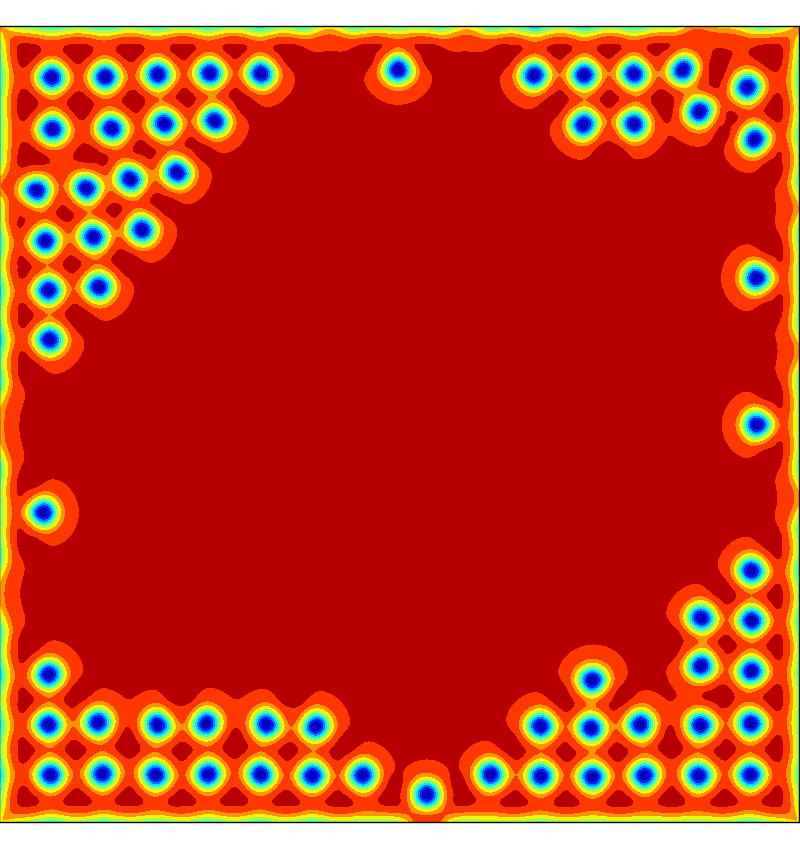}
\includegraphics[width=.19\linewidth]{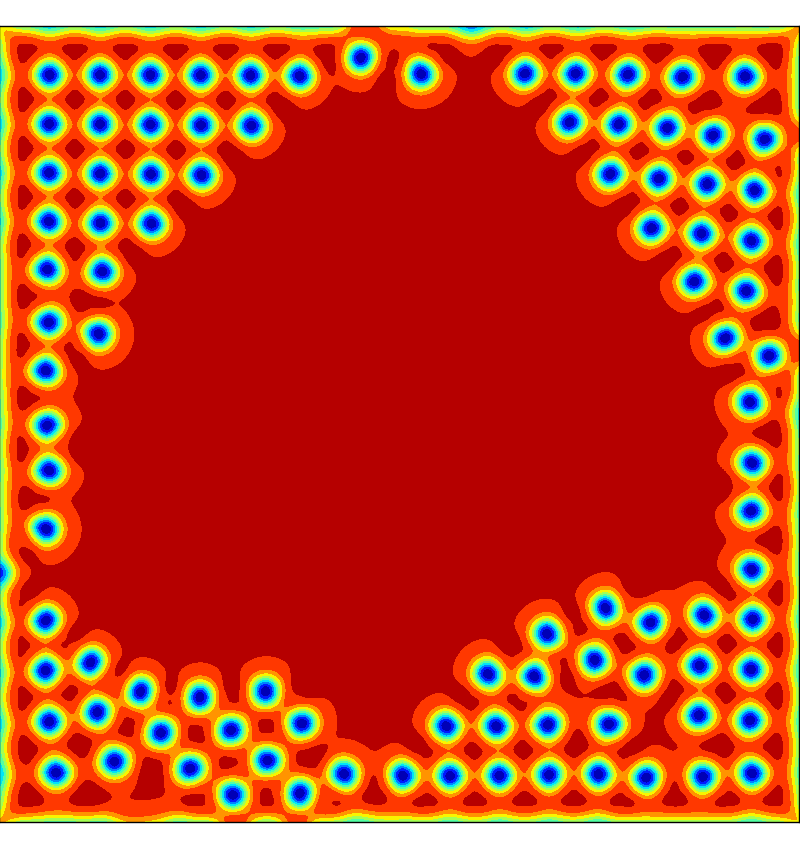}
\includegraphics[width=.19\linewidth]{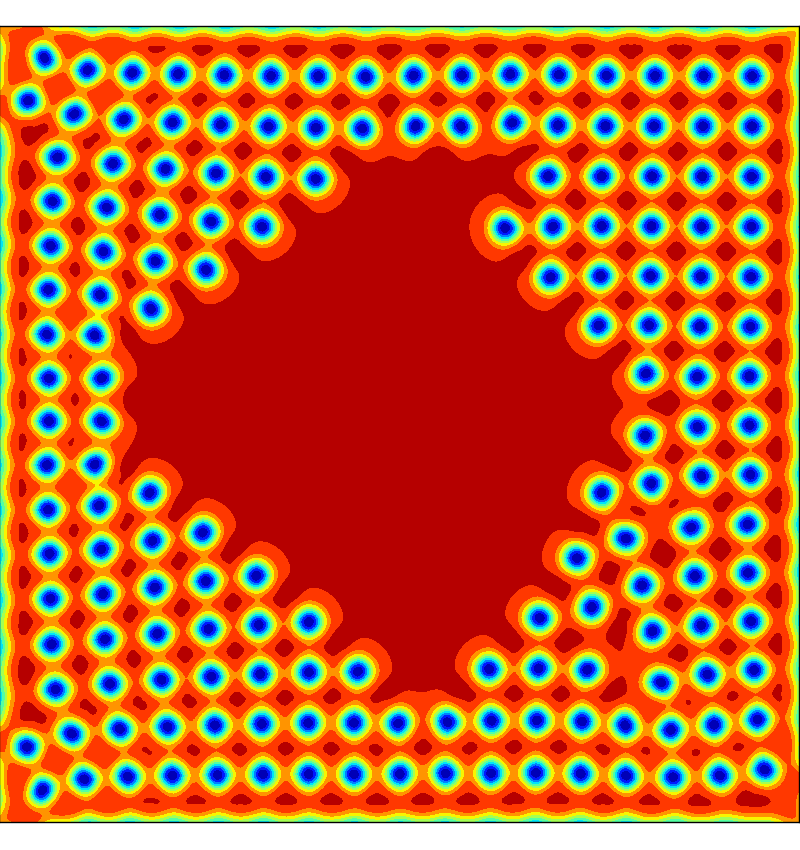}
\includegraphics[width=.19\linewidth]{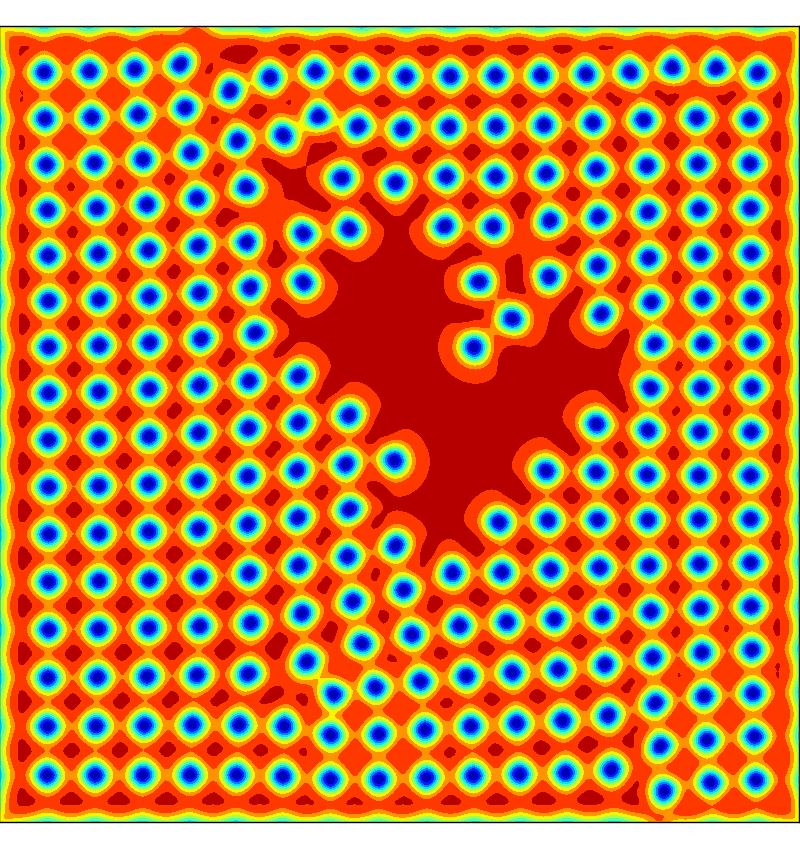}
\includegraphics[width=.19\linewidth]{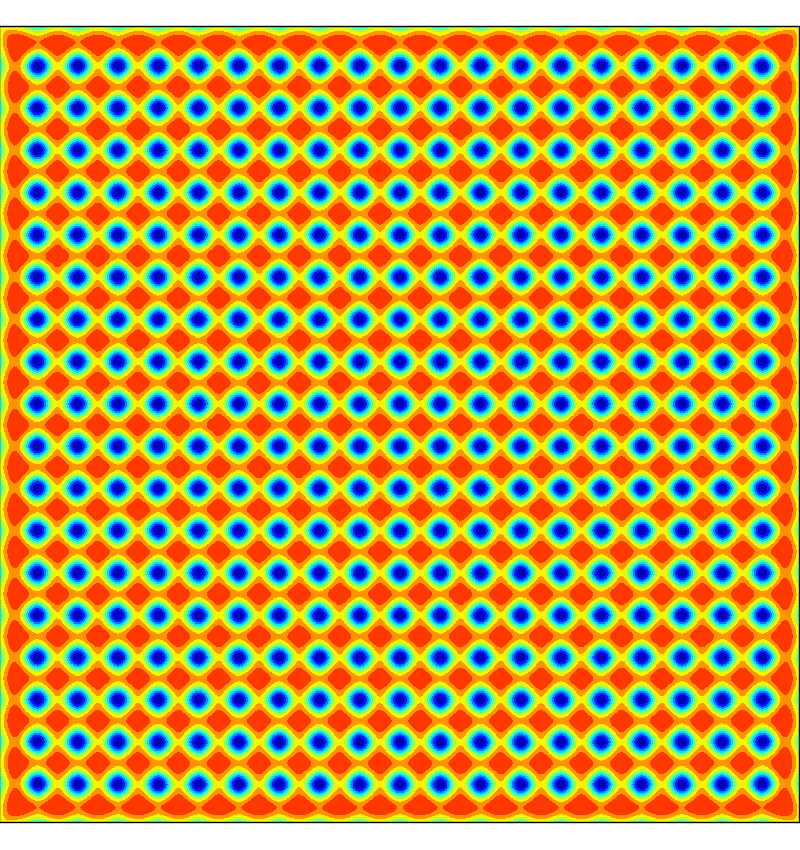}
\includegraphics[height=.201875\linewidth]{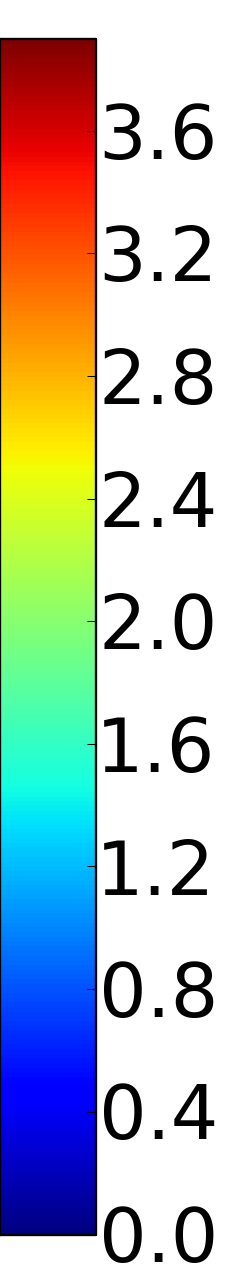}
\hss}
 \hbox to \linewidth{ \hss
\includegraphics[width=.19\linewidth]{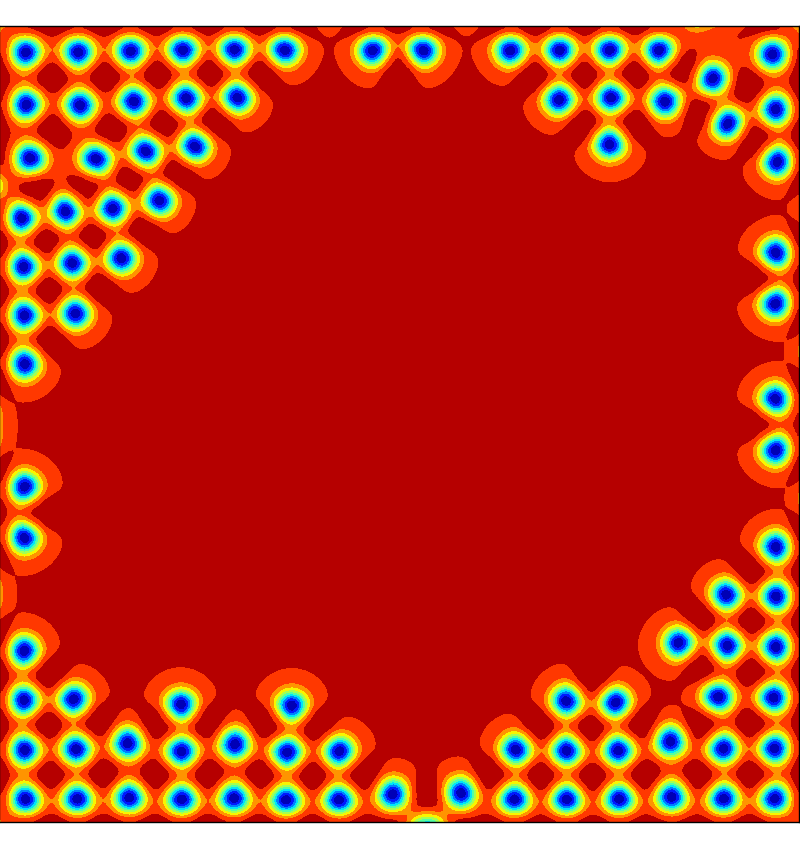}
\includegraphics[width=.19\linewidth]{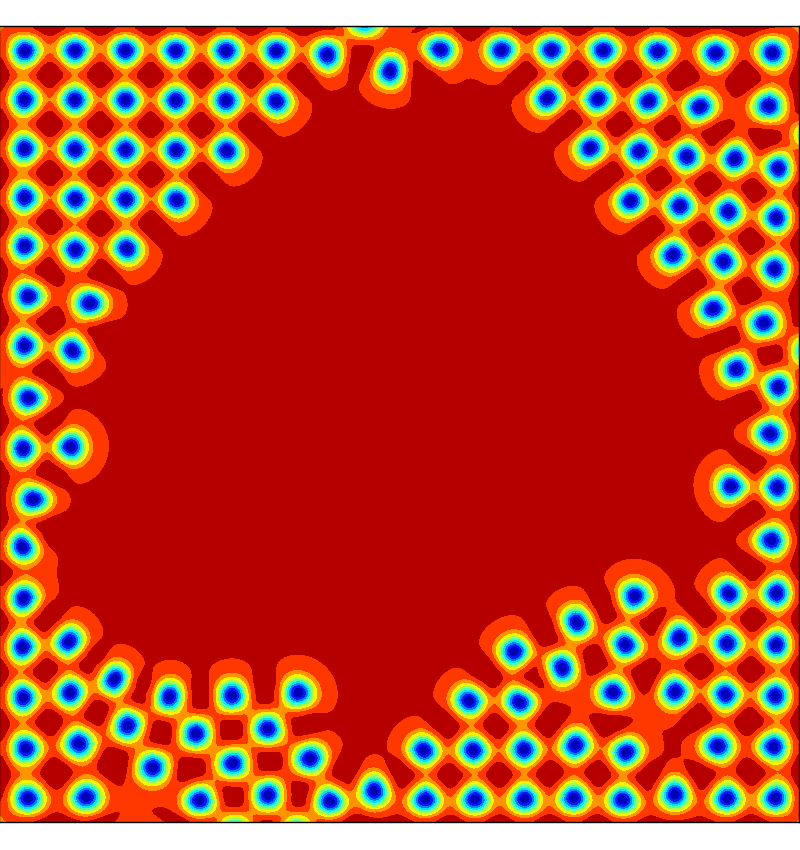}
\includegraphics[width=.19\linewidth]{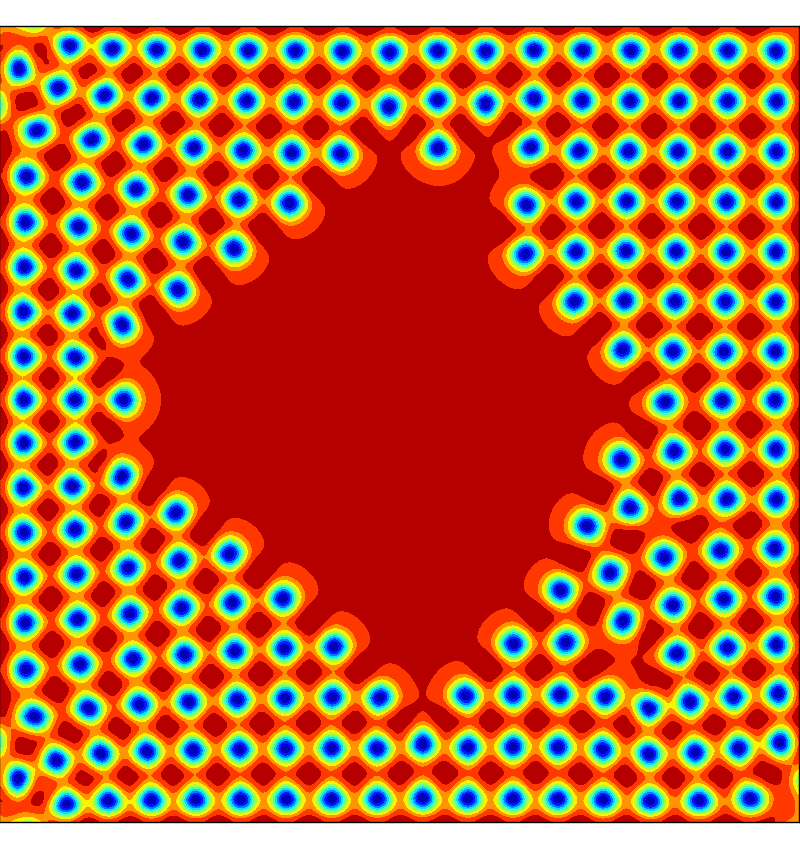}
\includegraphics[width=.19\linewidth]{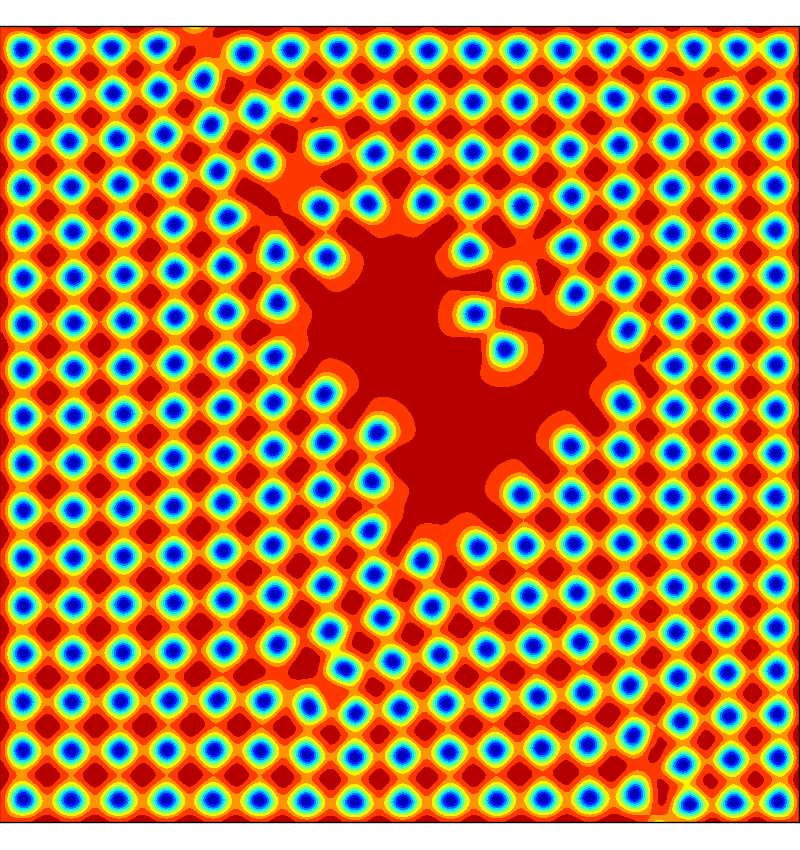}
\includegraphics[width=.19\linewidth]{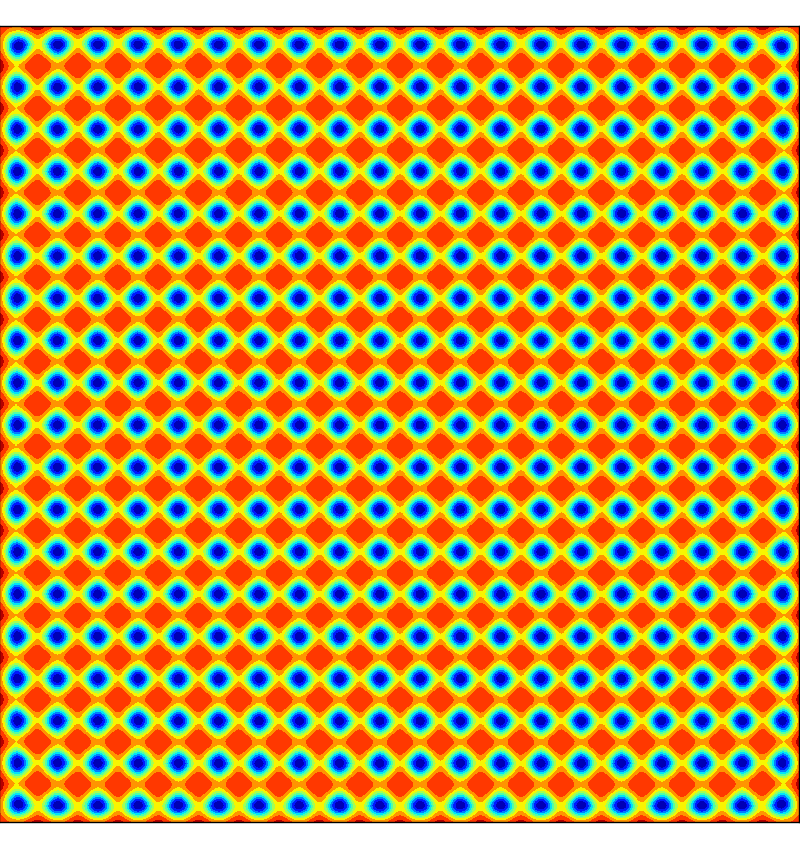}
\includegraphics[height=.201875\linewidth]{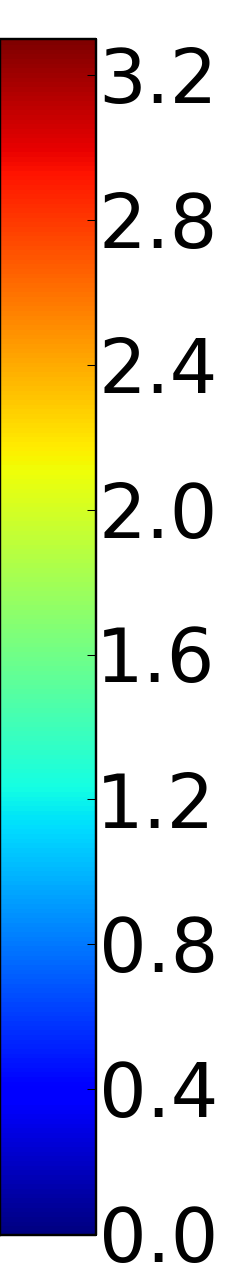}
\hss}
 \hbox to \linewidth{ \hss
\includegraphics[width=.19\linewidth]{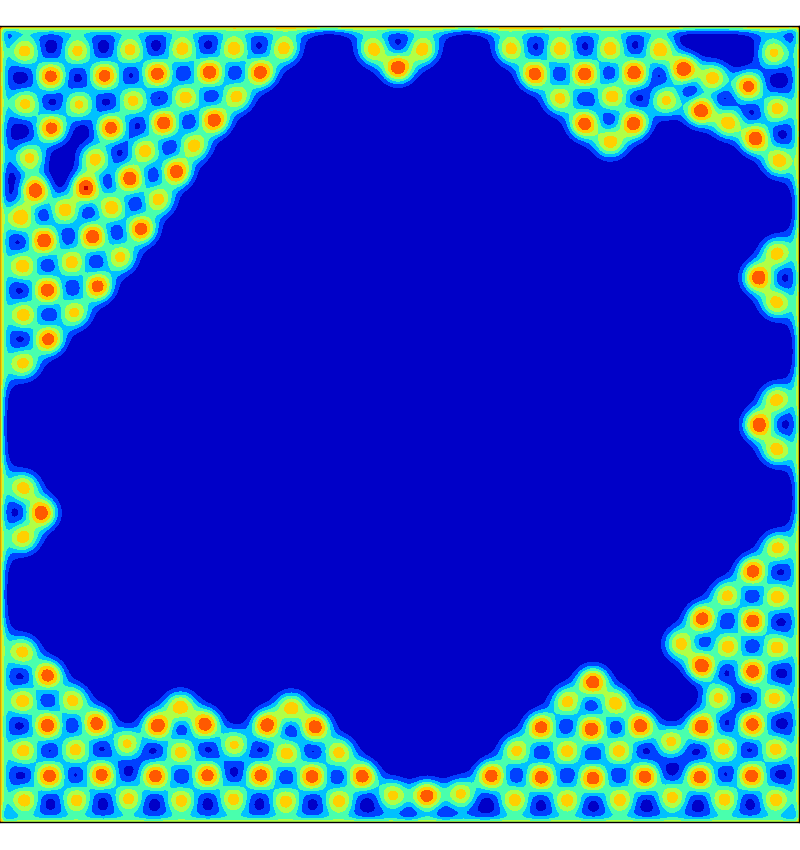}
\includegraphics[width=.19\linewidth]{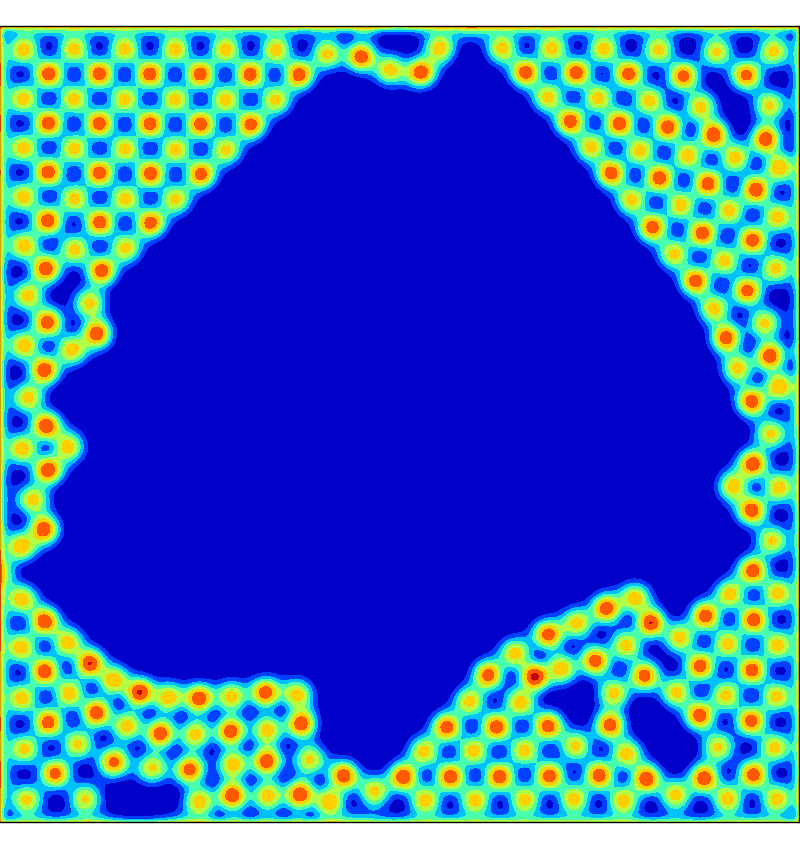}
\includegraphics[width=.19\linewidth]{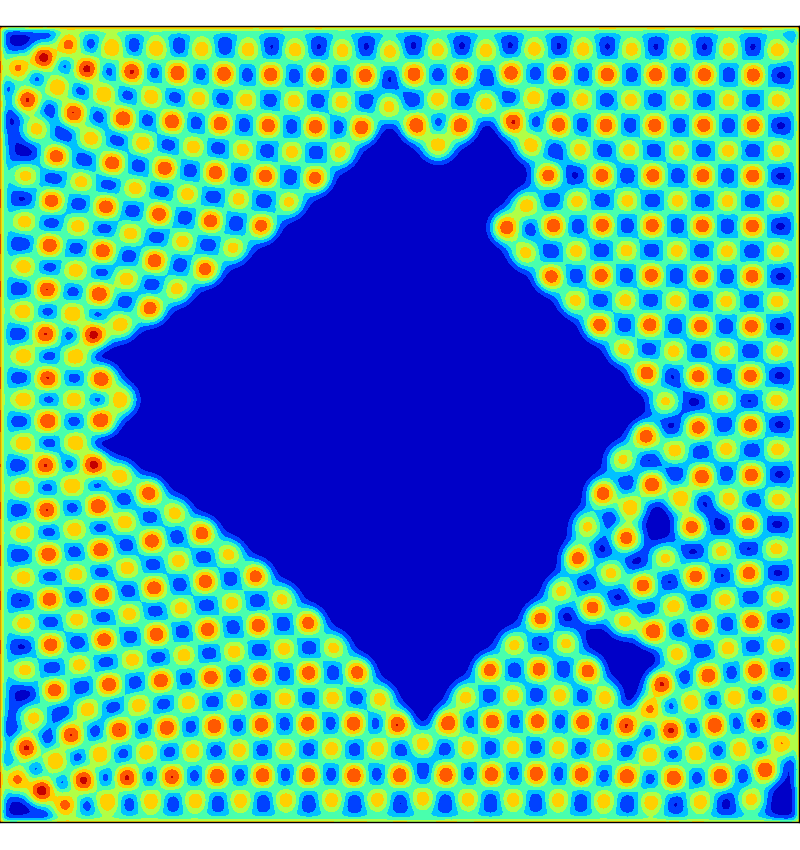}
\includegraphics[width=.19\linewidth]{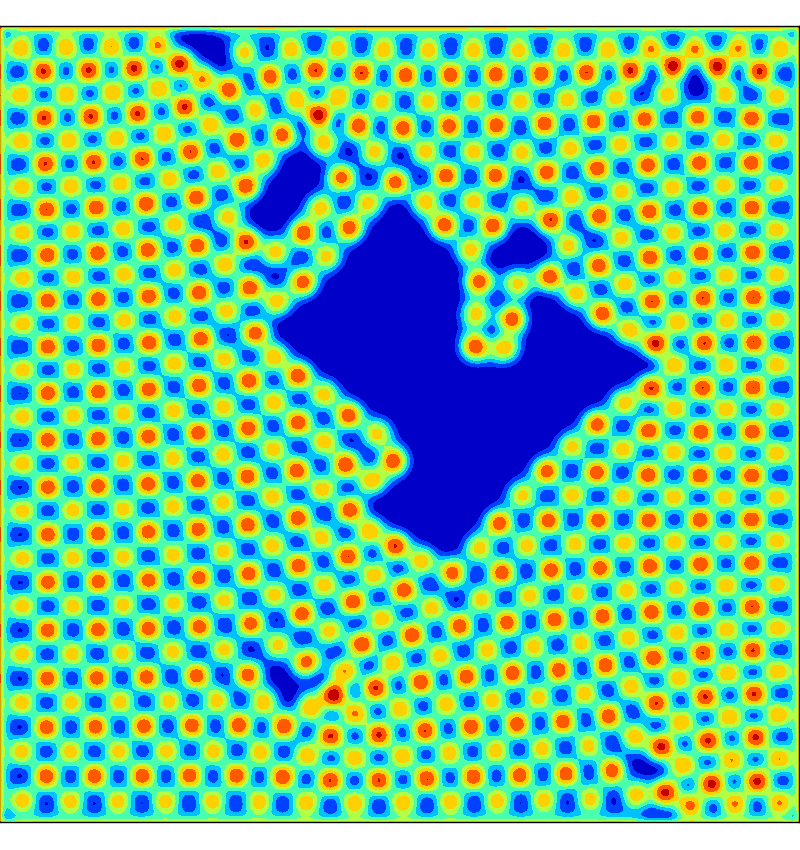}
\includegraphics[width=.19\linewidth]{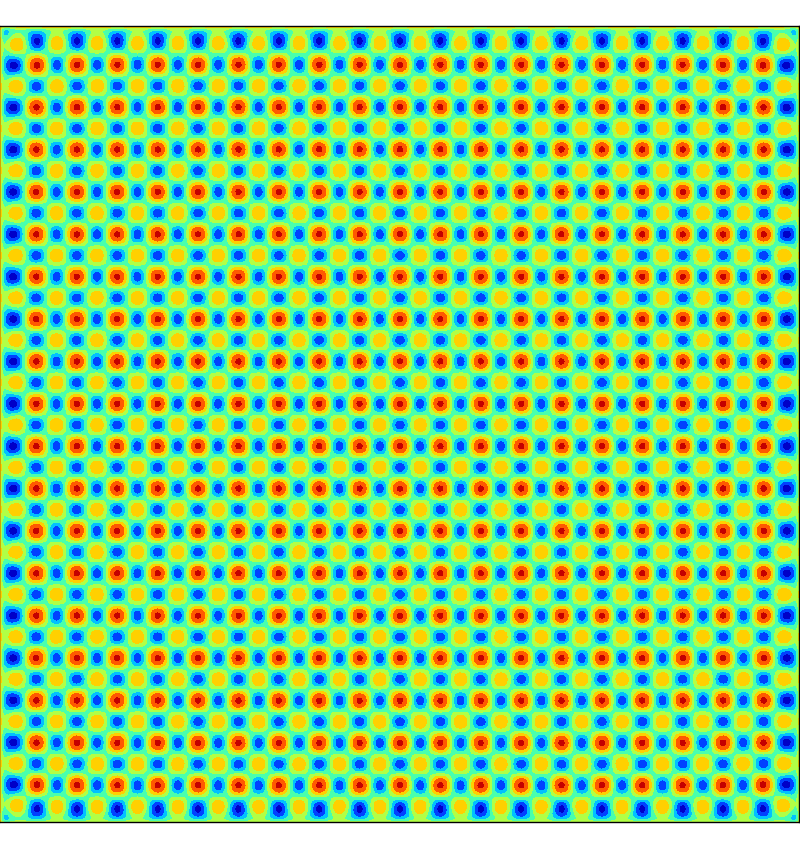}
\includegraphics[height=.201875\linewidth]{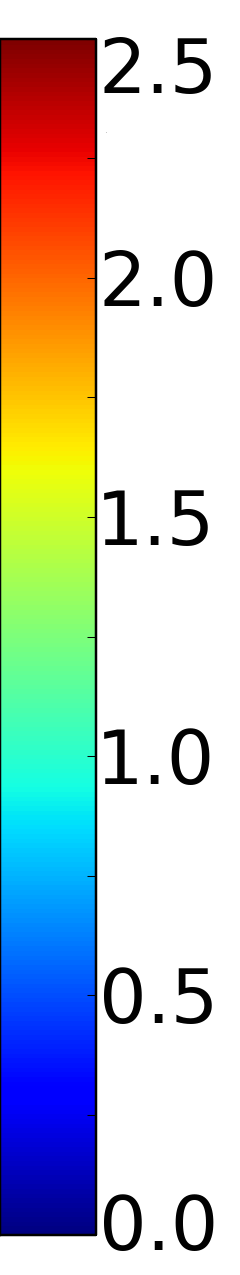}
\hss}
\caption{
(Color online) -- Sequences of the  Skyrmionic states in in the 
magnetization process of a finite sample in slowly increased 
magnetic flux. 
Corresponding values of 
the applied flux are respectively $96$, $129$, $201$, $258$ and $381$ 
(in the unit of the flux quantum). Parameters of the Ginzburg-Landau 
free energy are the same as in \Figref{Fig:Compact}.
First line shows $|\psi_1|^2$, second line $|\psi_2|^2$ while the third 
displays the magnetic field $\B$.
The peaks of different intensities in the magnetic field, correspond 
to vortices carrying different fractions of flux quantum. Note that 
there is a layer of half-Skyrmions near boundary. This is consistent 
with the thermodynamic stability of fractional vortices near boundaries 
as discussed in \cite{silaevf}.
Animations of the magnetization process are available as online 
Supplemental Material 
\cite{
[{See online Supplemental Material} ]
[{, for animations of the magnetization process. 
}]Supplementary}.
%
}
\label{magnetisation}
\end{figure*}
To investigate multi-quanta states we compute configurations carrying 
several flux quanta by energy minimization. First, as displayed in the 
first line in \Figref{Fig:Compact}, they can form compact `checkerboard' 
cluster. Unlike type-1.5 vortex clusters, where (composite) vortices can 
form cluster with inner triangular ordering \cite{Carlstrom.Garaud.ea:11,
Carlstrom.Garaud.ea:11a,Garaud.Agterberg.ea:12}, the dipolar-attraction 
driven structures have compact lattices with two interlaced square lattices 
\footnote{Note that formation of checkerboard square lattices for $p$-wave
superconductors near $H_{c2}$ were found in \cite{Agterberg:98,chung2}. 
Here we consider a different situation of vortex cluster formed due to 
attractive dipolar interactions.}.
Other kind of structures which we found for few vortex states are loop- 
and stripe- like structures. These are shown on \Figref{Fig:Selection} 
and details about these configurations are included 
Appendix~\ref{AppAdditional}. Some of these configurations are 
metastable local minima. The trend which we observed is that with 
increasing the drag coupling, multiple quanta configurations become 
more compact. Remarkably some of the vortex structures which we obtain 
are quite similar to those appearing in the easy-plane baby-Skyrme model 
consisting of the pseudo-spin $\bf n$ alone \cite{Jaykka.Speight:10}.
This similarity in structures is an interesting fact which could not be 
a priori expected because $\bf n$ represents only a part of the degrees 
of freedom of GL theory \Eqref{NLSM}, and does not account for all 
intervortex interaction forces. Moreover, at short length-scales, the 
GL model is certainly principally different from Skyrme model \cite{prb09}. Our 
observations demonstrate that at least in two dimensions there is a very 
close relationship between structure formation of topological defects in 
multicomponent superconductors and in pure baby-Skyrme models.
Besides that we find that structure formation  exhibit also complicated 
octagonal loop-like periodic structure as in the first line in 
\Figref{Fig:Honeycomb}. Their elementary cell carries $\Q=4$ flux quanta, 
and assumes octagonal geometry as a result of rotated underlying square 
fractional vortex structures.

Since the dipolar interactions are long-range they should dominate the 
tail of inter-skyrmion interactions. We therefore examine how much of 
the structure formation can be reproduced in the toy model of interacting 
point charges \Eqref{Interaction}. To this end we perform Monte Carlo (MC) 
simulations using the Metropolis algorithm with parallel tempering 
\cite{Earl.Deem:05}. 
Although the point-charge model does not perfectly capture all the 
underlying physics, it reproduces some aspects of the structures 
obtained beyond the London limit (see \Figref{Fig:Compact} and 
\Figref{Fig:Honeycomb}). 
Moreover, the MC approach allows to investigate how the ordering 
depends on temperature. As shown on \Figref{Fig:Compact}-(D) and 
\Figref{Fig:Honeycomb}-(D), thermal fluctuations can cause unbinding 
of the crystalline multi-quanta skyrmionic bound states held by dipolar 
forces. However fractional vortices are still paired and constitute 
well-defined skyrmions in higher temperature phases where there is no 
lattice structure. 

Finally we address the magnetization process of the skyrmionic state.
To this end we simulate the Gibbs free energy $\G=\F-\B\cdot {\bf H}$ 
of the system \Eqref{FreeEnergy}, on a finite domain in an increasing 
external field ${\bs H}=H\ez$. Here, finite differences are used instead 
of finite elements, and a quasi-Newton (BFGS) method instead of conjugate 
gradients. For details, see \cite{Palonen.Jaykka.ea:12} and 
Appendix~\ref{AppNumerics}. %
The magnetization process of the skyrmionic states is quite specific. 
It can be easily distinguished from other unconventional  magnetization 
processes such as those of chiral $p$-wave superconductors with multidomains 
\cite{machida}, entropically stabilized square lattices \cite{chung}, 
and type-1.5 superconductors \cite{Carlstrom.Garaud.ea:11,
Garaud.Agterberg.ea:12}. As shown on \Figref{magnetisation}, it is heavily 
influenced by the existence of dipolar forces. In these simulations we 
typically observed that multi-skyrmion domains bound by dipolar forces 
are formed near boundaries. These domains are attracted to boundaries by 
long range dipolar interaction with image charges. This crucially modifies 
Bean-Livingston barrier physics because dipolar attraction to the image 
``anti-skyrmions" has longer range than the  repulsion from the boundary 
due to surface Meissner current. These domains gradually fill the system 
until merging to form a (checkerboard) square lattice of fractional 
vortices. When the field is increased further the density of skyrmions 
in the square lattice grows. Importantly, during the magnetization process, 
the skyrmionic charge does not change in integer steps. When the condensates 
are not equivalent there is a layer of one kind of fractional vortices 
(or half-skyrmions) near boundaries as can be seen in \Figref{magnetisation}. 
This is in agreement with the thermodynamical stability of fractional 
vortices near boundaries demonstrated by Silaev, in the London limit 
without drag \cite{silaevf}.

In conclusion we investigated topological defects and magnetic response 
of $\Uone\times\Uone$ superconductors with dissipationless drag, beyond 
the commonly used London approximation. In contrast to the London limit, 
it requires a critical strength of dissipationless drag to form 
unconventional split vortex solutions. 
We demonstrated that split fractional vortices in this model have a well 
defined skyrmionic charge. We established that, when the model is 
$\Uone\times\Uone$ or softly-broken $\Uone\times\Uone$, the vortex 
lattice structure is dominated by the long-range dipolar inter-Skyrmion 
forces. This results in unconventional magnetic response in low fields 
which features lack of hexagonal vortex lattice and formation of a layer 
of square lattice growing inward from boundaries of the sample. This 
magnetization process can be easily identified for example in scanning 
SQUID measurements and discriminated from other models for $p$-wave 
superconductivity which by contrast predict hexagonal vortex lattices 
in low fields and square lattice in high fields.
It can also be straightforwardly distinguished from that of ordinary 
single-component type-II superconductors, or multicomponent type-1.5 
superconductors or chiral $p$-wave multi-domain superconductors. 
For example the magnetic behavior of the putative triplet superconductor 
Sr$_2$RuO$_4$ is nontrivial, featuring phase separation \cite{ref4,ref2, 
ref1,curran,ref3}. However since square vortex lattices were observed 
only at elevated fields and no boundary vortex states were reported, 
it is inconsistent with models which have long-range skyrmionic forces.

We thank Johan Carlstr\"om for discussions.
This work is supported by the Swedish Research Council, by the Knut and 
Alice Wallenberg Foundation through the Royal Swedish Academy of Sciences 
fellowship and by NSF CAREER Award No. DMR-0955902.
The computations were performed on resources provided by the Swedish 
National Infrastructure for Computing (SNIC) at National Supercomputer Center 
at Linkoping, Sweden.

\appendix
\setcounter{section}{0}
\setcounter{equation}{0}
\renewcommand{\theequation}{\Alph{section}.\arabic{equation}}

\section{Details of theoretical framework}
\label{AppTheory}

In two-component superconductors, the elementary topological excitations 
are fractional vortices. These are field configurations having a $2\pi$ 
phase winding only in one phase (\eg $\varphi_1$ has 
$\oint\Grad\varphi_1=2\pi$ winding while $\oint\Grad\varphi_2=0$). 
The physics of fractional vortices, as well as the role of the intercomponent 
dissipationless drag can be enlightened by rewriting the theory in terms 
of \emph{charged} and \emph{neutral} modes. Here we derive the interaction 
between fractional vortices, for a two-component system. In particular 
this shows how, in the London limit, fractional vortices can be treated 
as point particles with Coulomb and Yukawa interactions. The Ginzburg-Landau 
free energy functional reads ss
\SubAlign{AppFreeEnergy}{
 \F&= \frac{1}{2}(\bs \nabla \times \A)^2
  +\sum_{a}\frac{1}{2}|\D\psi_a|^2   					\label{AppMagneticEnergy}\\
&+\sum_{a}\alpha_a|\psi_a|^2+\frac{1}{2}\beta_a|\psi_a|^4 	\label{AppPotential1} \\
&+\frac{1}{2}\gammaot|\psi_1|^2|\psi_2|^2			\label{AppPotential2}  \\
&+\frac{\nuot}{2}
\left|	\Im(\psi_1^*\D\psi_1)+\Im(\psi_2^*\D\psi_2)\right|^2	\label{AppAndreev-Bashkin}\,.
}
Note that for completeness we added bi-quadratic density coupling 
\Eqref{AppPotential2} to the potential energy,.
Obviously, its effect is to enforce core splitting of fractional 
(when $\gammaot>0$) vortices. Since we mostly focus on the role of 
the drag interaction, the bi-quadratic density term is introduced here 
for sake of completeness rather than an essential ingredient of the 
physics we discuss.

\subsection{Parametrization of the intercomponent drag and its London limit}

Traditionally, the intercomponent current-current interaction is 
parametrized as the scalar product of supercurrents of two components
$\F_d \propto \J_1 \cdot \J_2$. Beyond the London limit, such a term reads 
explicitly $\F_d\propto \Im(\psioc\D\psio)\cdot\Im(\psitc\D\psit)$. This 
term is fourth order in the order parameters densities and second order 
in their derivatives, moreover it is not positively defined. This leads 
to an unphysical instability: by creating strong counter-directed currents
and increasing density, in a minimal GL model with such a term,  makes free 
energy negative and unbounded from below. Thus this term should come with 
other high-power terms consistent with the symmetry, which make the total 
free energy positively defined. The precise form of these terms is not 
principally important for the purpose of this work, so we choose to use 
\Eqref{AppAndreev-Bashkin}, which is obviously positive. However one 
should also make sure that this term has the proper London limit. 
There, the free energy functional \Eqref{AppFreeEnergy} reads as
\SubAlign{AppFreeEnergyLL}{
 \F&= \frac{1}{2}(\bs \nabla \times \A)^2
  +\sum_{a}\frac{1}{2}|\D\psi_a|^2   	\label{AppMagneticEnergyLL}\\
&+\frac{\nuot}{2}\left|	\Im(\psi_1^*\D\psi_1)
+\Im(\psi_2^*\D\psi_2)\right|^2		\label{AppAndreev-BashkinLL}\,.
}
Since the densities are constant, the covariant derivative 
reads as
$
   \D\psia=i|\psia|(\Grad\varphi_a+e\A)\Exp{i\varphi_a}
$
and thus, expanding the drag term \Eqref{AppAndreev-BashkinLL} 
and collecting various orders, the free energy assumes the
form typically used for discussing the problem in the London limit
\SubAlign{AppSF}{
\F&= \frac{1}{2}(\Curl\A)^2 
	+\sum_{a=1,2}\frac{1}{2}\rho_{aa}(\Grad\varphi_a+ e\A)^2  \label{AppSF1}\\
&+\rho_{d}(\Grad\varphi_1+ e{\bs A})\cdot 
   (\Grad\varphi_2+ e\A)  \,.\label{AppSF2}
}
Where the prefactors are 
\Align{AppSFpref}{
   \rho_{aa}&=|\psi_a|^2(1+\nuot|\psi_a|^2) \nonumber \\
   \rho_{d}&=\nuot|\psi_1|^2|\psi_2|^2 \,.
}
The term \Eqref{AppSF2} is the scalar product of the 
supercurrents of two components. Thus our parametrization 
\Eqref{AppAndreev-Bashkin} of intercomponent current-current 
interaction has the conventional London limit.

\subsection{Derivation of neutral and charged modes}

To understand the role of the fundamental excitations (\ie fractional 
vortices), the Ginzburg-Landau free energy \Eqref{AppFreeEnergy}
can be rewritten into \emph{charged} and \emph{neutral} modes 
by expanding the kinetic term \Eqref{AppMagneticEnergy} and the drag term
  \Eqref{AppAndreev-Bashkin}
\SubAlign{AppGLRewritten1}{
   \F&= \frac{1}{2}(\Curl \A)^2 + \frac{\J^2}{2e^2w\varrho^2} \label{AppChargedMode}\\
 &+\sum_{a}\frac{1}{2}(\Grad|\psi_a|)^2
+\alpha_a|\psi_a|^2+\frac{\beta_a}{2}|\psi_a|^4 \label{AppHiggsMode0} \\
&  +\gammaot|\psi_1|^2|\psi_2|^2			\label{AppHiggsMode} \\
&+\frac{|\psi_1|^2|\psi_2|^2}{2\varrho^2}(\Grad\varphiot)^2 \
\label{AppNeutralMode}	\,.
}
Here $\varphiot\equiv\varphi_2-\varphi_1$ is the phase 
difference and 
\Equation{AppNotations}{
w=1+\nuot\varrho^2~~~~\text{and}~~~~
\varrho^2=\sum_a|\psi_a|^2
\,.
}
The supercurrent defined from the Amp\`ere's equation $\Curl\B+\J=0$, 
reads as
\SubAlign{AppCurrents3}{
  \J/e&= ew\varrho^2\bs A+\sum_{a}|\psi_a|^2\bs \nabla\varphi_a  \label{AppCurrents3a}\\
&+\nuot( |\psi_1|^2+ |\psi_2|^2) 
(|\psi_1|^2\bs\nabla\varphi_1+|\psi_2|^2\bs\nabla\varphi_2) \nonumber \\
      &=ew\varrho^2\bs A+w\sum_{a}|\psi_a|^2\bs\nabla\varphi_a
\label{AppCurrents3b} \,,
}
while the supercurrent associated with a given condensate reads as
\Align{AppCurrents4}{
     \J_a  &=e\Im(\psiac\D\psia)\left(1+\nuot|\psia|^2\right)	\nonumber \\
	    &+|\psia|^2\nuot e\Im(\psibc\D\psib)	\,, 
}
with the band index $b\neq a$. The term on the second line is the 
current of the component $a$ induced (dragged) by the component $b$.
Assuming phase winding in all components and since far away from a 
vortex $\J$ decays exponentially, the magnetic flux reads as
\Align{AppFlux}{
   \Phi&=\int\B\dd S=\oint \A\dd\ell 	\nonumber\\
	 &=\frac{1}{e^2w\varrho^2}
	    \oint \left(\J-ew\sum_{a}|\psi_a|^2\bs\nabla\varphi_a\right)\dd\ell \nonumber \\
	 &=\Phi_0\sum_{a}\frac{|\psi_a|^2}{\varrho^2} \,,
}
where $\Phi_0=2\pi/e$ is the flux quantum and the closed path integration 
is done so that the flux is positive. The fraction of flux 
$|\psi_a|^2\Phi_0/\varrho^2$ carried is the same as that of two-component 
superconductors without drag \cite{frac}. The London limit, assumes that 
$|\psi_a| =\mathrm{const}$ everywhere in space except small vortex core sharp 
cut-off. The expression \Eqref{AppGLRewritten1} thus further simplifies
\SubAlign{AppGLRewritten2}{
   \F&= \frac{1}{2}\left(\B^2 + \frac{1}{e^2w\varrho^2} 
   |\Curl\B|^2\right)\label{AppChargedMode2}\\
 &+\frac{|\psi_1|^2|\psi_2|^2}{2\varrho^2}(\Grad\varphiot)^2 
\label{AppNeutralMode2}	\,,
}
where the Amp\`ere's law has been used to replace the 
current in \Eqref{AppChargedMode2}.
The interaction energy of two non-overlapping fractional 
vortices can be approximated in this London limit by considering 
\emph{charged} \Eqref{AppChargedMode2} and \emph{neutral} 
modes \Eqref{AppNeutralMode2}, separately. 
With the identity 
\Equation{AppIdentity}{
|\Curl\B|^2=\B\cdot\Curl\Curl\B-\Div(\B\times\Curl\B)\,,
}
the energy of the \emph{charged} sector 
\Eqref{AppChargedMode2} finally reads 
\Equation{AppChargedMode3}{ 
   F_{\mbox{\tiny Charged}}=\int \frac{\B}{2}\left( 
\B+\frac{1}{e^2w\varrho^2}\Curl\Curl\B
\right)\,.
}
The London equation for a (point-like) vortex placed 
at $\x_a$ and carrying a flux $\Phi_a$ is 
\Equation{AppAmpere}{
   \frac{1}{e^2w\varrho^2}\Curl\Curl\B+\B=\Phi_a\delta(\x-\x_a)\,,
}
and its solution is 
\Equation{AppBfield}{
   \B_a(\x)=\frac{\Phi_ae^2w\varrho^2}{2\pi}
   K_0\left(\frac{|\x-\x_a|}{\lambda}\right)\,.
 }
Here the London penetration length is $\lambda=\frac{1}{e\sqrt{w\varrho^2}}$
and $K_0$ is the modified Bessel of second kind. For two 
vortices located at $\x_a$ and $\x_b$, and carrying fluxes 
$\Phi_a$ and $\Phi_b$, the source term in London equation 
reads $\Phi_a\delta(\x-\x_a)+\Phi_b\delta(\x-\x_b)$ and the 
magnetic field is the superposition of two contributions 
$\B(\x)=\B_a(\x)+\B_b(\x)$. Thus  
\Align{AppAmpere2}{
    F_{\mbox{\tiny Charged}}&=\int\frac{1}{2}
   (\B_a+\B_b )(\Phi_a\delta(\x-\x_a)+\Phi_b\delta(\x-\x_b)) \nonumber \\
   &=   \frac{\Phi_a\Phi_be^2w\varrho^2}{2\pi}K_0\left(\frac{|\x_2-\x_1|}{\lambda}\right)
   +E_{va}+E_{vb} \,,
}
and $E_{va}\equiv\int\B_a(\x_a)\Phi_a/2$ denotes the 
(self-)energy of the vortex $a$. Finally, the interaction 
energy of two vortices in components $a,b$ reads 
\Equation{AppChargedMode4}{ 
   E^{(int),\mbox{\tiny Charged}}_{ab}=\frac{2\pi w|\psi_a|^2|\psi_b|^2}{\varrho^2}
   K_0\left(\frac{|\x_a-\x_b|}{\lambda}\right)\,.
}
The interaction of the charged sector is thus a Yukawa-like interaction 
given by the modified Bessel function. If we do not consider anti-vortices 
it is  always positive (for any $a,b$), then it gives repulsive interaction 
between any kind of fractional vortices.
\begin{figure*}[!htb]
 \hbox to \linewidth{ \hss
 \includegraphics[width=.70\linewidth]{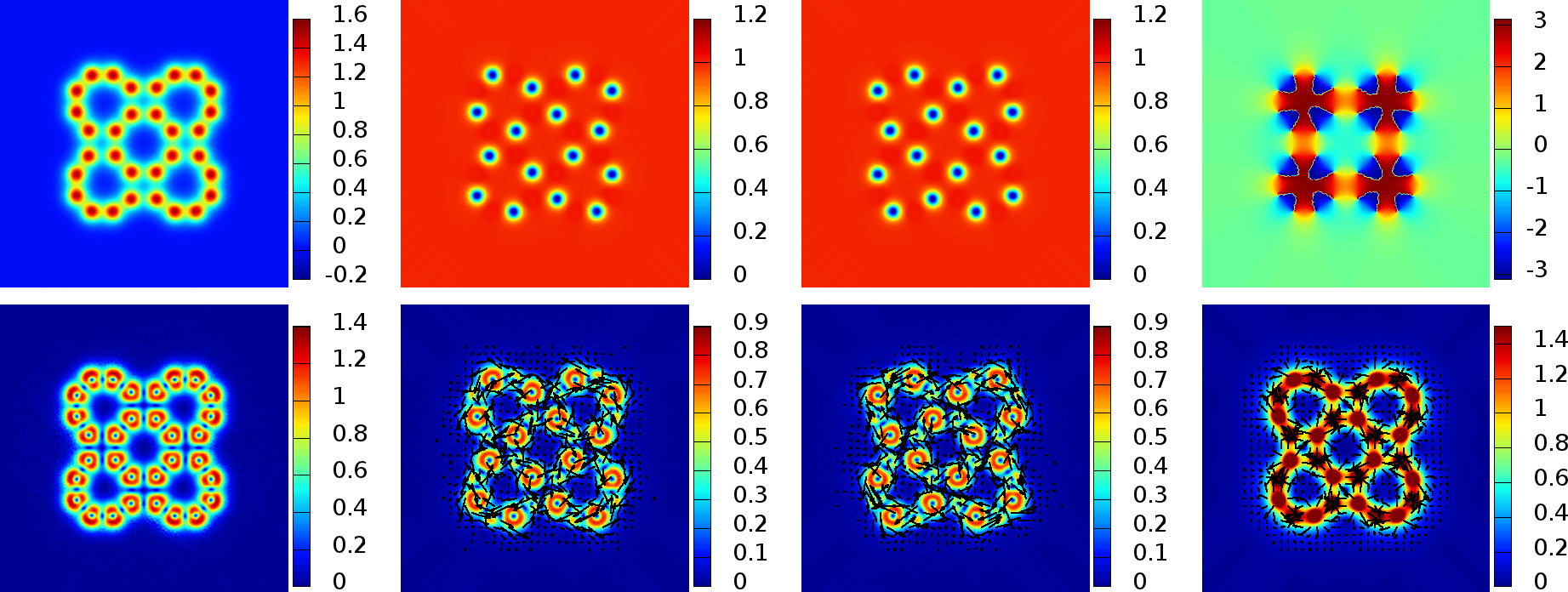}
\hspace{0.1cm}
 \includegraphics[width=.265\linewidth]{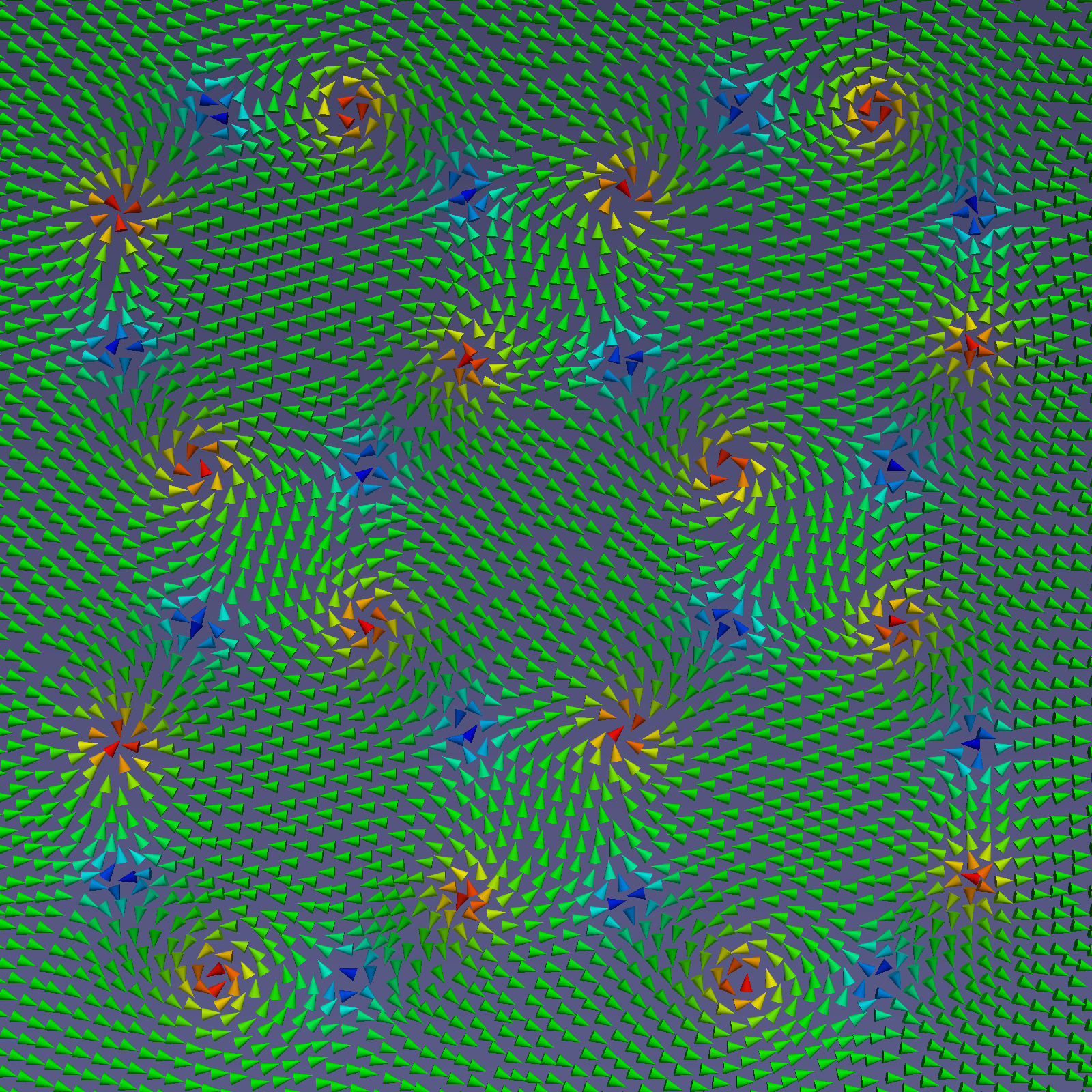}
 \hss}
\caption{
(Color online) -- Octagon-like structure carrying $\Q=16$ 
flux quanta. The elementary cell here is a$\Q=4$ skyrmion.
The parameters are $(\alpha_a,\beta_a)=(-5.0,1.0)$ 
with $e=0.6$ and Andreev--Bashkin coupling is 
$\nuot=2.0$, while bi-quadratic coupling 
vanish $\gammaot=0$.
Displayed quantities are respectively the magnetic field $\B$, $|\psi_1|^2$, 
$|\psi_2|^2$ and the phase difference $\varphiot\equiv\varphi_2-\varphi_1$, 
on the first line. On the second line, $\J$, $\J_1$, $\J_2$ and 
${\bf z}\times\Grad\varphiot$.
The rightmost panel shows the normalized projection of ${\bf n}$ 
onto the plane, while colors give the magnitude of $n_z$. Blue corresponds 
to the south pole (-1) while red is the north pole (+1) of the target sphere 
$S^2$. 
}
\label{Fig:Honeycomb3}
\end{figure*}
On the other hand, the interaction through the \emph{neutral} sector 
is logarithmic. It is attractive (\resp repulsive) for fractional 
vortices of the different (\resp same) kind. The energy associated 
with the \emph{neutral} mode \Eqref{AppNeutralMode2} reads 
\Equation{AppNeutralMode3}{ 
   F_{\mbox{\tiny Neutral}}=
\frac{|\psi_1|^2|\psi_2|^2}{2\varrho^2}\int(\Grad\varphiot)^2\,.
}
A phase winding around some singularity located at the 
point $\x_a$, is (at sufficiently large distance) well approximated 
by $\varphi_a=\theta$. Thus 
\Equation{AppGradPhi}{
\Grad\varphi_a=\frac{\et}{|\x-\x_a|}=\bs z\times\Grad\ln|\x-\x_a|\,.
}
To evaluate the interaction between fractional vortices in different 
condensates and respectively located at $\x_a$ and $\x_b$, the neutral 
sector is expanded 
\Align{AppNeutralModeLike1}{ 
   F_{\mbox{\tiny Neutral}}=
\frac{|\psi_1|^2|\psi_2|^2}{2\varrho^2}\int&
(\Grad\varphi_a)^2+(\Grad\varphi_b)^2 \nonumber \\
&-2\Grad\varphi_a\cdot\Grad\varphi_b
\,.
}
Thus the interacting part reads
\Align{AppNeutralModeLike2}{ 
   E^{(int),\mbox{\tiny Neutral}}_{ab}&=
-\frac{|\psi_1|^2|\psi_2|^2}{\varrho^2}\int\Grad\varphi_a\cdot\Grad\varphi_b \nonumber \\
&=\frac{|\psi_1|^2|\psi_2|^2}{\varrho^2}\int\varphi_a\Delta\varphi_b\nonumber \\
&=\frac{|\psi_1|^2|\psi_2|^2}{\varrho^2}
\int\ln|\x-\x_a|\delta(|\x-\x_b|)\nonumber \\
&=2\pi\frac{|\psi_1|^2|\psi_2|^2}{\varrho^2}\ln|\x_b-\x_a|
\,.
}
Similarly, the interaction between two vortices in the 
same condensate $a$ is computed by requiring that 
the phase is the sum of the individual phases 
$\varphi_a=\varphi_a^\oo+\varphi_a^\ot$, while 
$\varphi_b=0$. Then the interaction reads
\Equation{AppNeutralModeUnLike}{ 
   E^{(int),\mbox{\tiny Neutral}}_{aa}=
-2\pi\frac{|\psi_1|^2|\psi_2|^2}{\varrho^2}\ln|\x_a^\ot-\x_a^\oo|
\,.
}

To summarize, the interaction of vortices in different 
condensates is then
\Equation{AppInteractionUnlike}{
\frac{E^{(int)}_{12}}{2\pi}=\frac{|\psi_1|^2|\psi_2|^2}{\varrho^2}
    \Big(\ln \frac{r}{R} +wK_0\left(\frac{r}{\lambda}\right)\Big)\,,
}
while interactions of vortices of similar condensates are 
\Equation{AppInteractionLike}{
\frac{E^{(int)}_{aa}}{2\pi}= 
-\frac{|\psi_1|^2|\psi_2|^2}{\varrho^2} \ln\frac{r}{R} 
+\frac{w|\psi_a|^4}{\varrho^2} K_0\left(\frac{r}{\lambda}\right)\,,
}
with $r\equiv|\x_a-\x_b|$ and $R$ the sample size.
Equations \Eqref{AppInteractionUnlike} and  \Eqref{AppInteractionLike}
give the different interactions between fractional vortices.
Finally, choosing the energy scale to be $2\pi|\psi_1|^2|\psi_2|^2/\varrho^2$  
and defining the parameters $m$ 
and $w$ as
\Equation{AppInteraction_param}{
w =1+\nuot\varrho^2=1+\nuot (|\psi_1|^2+|\psi_2|^2)\,,
~~	m =\frac{|\psi_1|^2}{|\psi_2|^2} \,.	
}
The interaction between fractional vortices in the various 
condensates reads
\Align{AppInteraction}{
E_{11}&=\ln\frac{R}{r}+wmK_0\left(\frac{r}{\lambda}\right) 			\,,\nonumber \\
E_{22}&=\ln\frac{R}{r}+\frac{w}{m}K_0\left(\frac{r}{\lambda}\right)	\,,\nonumber \\
E_{12}&=-\ln\frac{R}{r}+wK_0\left(\frac{r}{\lambda}\right) \,.
}
Thus vortex matter in the London limit of a two-component
superconductor with intercomponent drag interaction is described 
by a 3-parameter family $(m,w,R)$.

\subsection{Mapping to an easy-plane non-linear 
\texorpdfstring{$\sigma$}{sigma}-model}

The bound state of well separated fractional vortices is a \emph{Skyrmion}. 
This follows from mapping the two-component model \Eqref{AppFreeEnergy} to 
an easy-plane non-linear $\sigma$-model \cite{bfn,prb09}. 
There, the pseudo-spin unit vector $\bf n$ is the projection of 
superconducting condensates on spin-$1/2$ Pauli matrices $\bs\sigma$: 
\Equation{Projection}{
 {\bf n}\equiv (n_x,n_y,n_z)
 =\frac{\Psi^\dagger\bs \sigma\Psi}{\Psi^\dagger\Psi}\,,
 ~~\text{where}~~
 \Psi^\dagger=(\psi_1^*,\psi_2^*)\, .
}
The following identity is useful to rewrite the free energy 
\Eqref{AppFreeEnergy} in terms of the pseudo-spin $\bf n$, total density 
$\varrho$ and gauge invariant current $\J$
\Align{AppNLS-Identity1}{
\frac{\varrho^2}{4}\partial_kn_a\partial_kn_a+(\Grad\varrho)^2=& 
\frac{|\psi_1|^2|\psi_2|^2}{\varrho^2}(\Grad\varphiot)^2 \nonumber \\
&+\sum_a(\Grad|\psi_a|)^2 \,,
}
where summation on repeated indices is implied. Using the definition 
of the current \Eqref{AppCurrents3} and noting that 
\Equation{AppNLS-Identity2}{4
\varepsilon_{ijk}\partial_i\left(
\sum_a\frac{|\psi_a|^2}{\varrho^2}\partial_j\varphi_a\right) = 
\varepsilon_{ijk}\varepsilon_{abc}n_a\partial_in_b\partial_jn_c ,
}
where $\varepsilon$ is the Levi-Civita symbol, the magnetic field reads 
\Equation{AppNLS-Identity3}{
B_k=\frac{1}{e}\varepsilon_{ijk}\left(\partial_i\left( \frac{J_j}{ew\varrho^2} \right)
-\frac{1}{4}\varepsilon_{abc}n_a\partial_in_b\partial_jn_c \right)\,,
}
and the free energy \Eqref{AppGLRewritten1} can be written as
\Align{AppNLSM}{
 \F&= \frac{1}{2}(\Grad\varrho)^2+\frac{\varrho^2}{8}\partial_k n_a\partial_k n_a
  	+\frac{\J^2}{2e^2w\varrho^2}+V(\varrho,n_z)	\nonumber \\
+&\frac{1}{2e^2}\left[\varepsilon_{ijk}\left(
\partial_i \left(\frac{J_j}{ew\varrho^2}\right)
	-\frac{1}{4}\varepsilon_{abc}n_a\partial_i n_b\partial_j n_c \right)\right]^2
 \,,
}
where $V(\varrho,n_z)$ stands for the potential terms \Eqref{AppPotential1} 
and \Eqref{AppPotential2}. The easy plane potential explicitly reads
\Equation{AppNLSMPotential}{
V(\varrho,n_z)=\frac{\varrho^2}{2}(a_1+a_2n_z)
+\frac{\varrho^4}{4}(b_1+2b_2n_z+b_3n_z^2)\,,
}
with the coefficients
\Align{AppNLSMPotentialCoeff}{
b_1&=\frac{\beta_1+\beta_2+\gamma}{2}\,,~~~
b_2=\frac{\beta_1-\beta_2}{2}\,,~~~
b_3=\frac{\beta_1+\beta_2-\gamma}{2}\,,\nonumber \\
a_1&=\alpha_1+\alpha_2\,,~~~
a_2=\alpha_1-\alpha_2\,.
}
The pseudo-spin is a map from the one-point compactification of the 
plane ($\Real^2\simeq S^2 $) to the two-sphere target space spanned 
by $\bf n$. That is ${\bf n}: S^2\to S^2$, classified by the homotopy 
class $\pi_2(S^2)\in\Relative$, thus defining the integer valued 
topological (skyrmionic) charge  
\Equation{Charge}{
   \Q({\bf n})=\frac{1}{4\pi} \int_{\Real^2}
   {\bf n}\cdot\partial_x {\bf n}\times \partial_y {\bf n}\,\,
  \dd x \dd y \,.
}
Ordinary (composite) vortices with a single core $\Psi=0$, have
$\Q=0$. Core split vortices, on the other hand, have non-trivial 
skyrmionic charge $\Q=N$ (with $N$ coincides with the number of 
carried flux quanta). The calculated pseudo-spin texture of $\bf n$ 
is shown on the rightmost panel in \Figref{Fig:Honeycomb3}. Numerically 
calculated topological charge was found to be integer (with a 
negligible error of order $10^{-5}$).
It is worth emphasizing that the topological charge \Eqref{Charge} is 
an integer, when integrated over the infinite plane $\Real^2$, or at 
least an large enough domain $\Omega\subset\Real^2$. By large enough, 
we understand that the fields should have recovered their ground state 
values at the boundary. Then the skyrmions shall not interact with the 
boundary. When the Skyrmion's size is comparable with the size of the 
integration domain, truncation error appear and $\Q$ is no more integer. 
Moreover when simulating a finite sample in applied field, in general 
the skyrmionic topological charge $\Q$ will not be integer. This is 
because in general there are states where only a part of the Skyrmion 
texture enters the sample.

\begin{figure*}[!htb]
 \hbox to \linewidth{ \hss
 \includegraphics[width=0.8\linewidth]{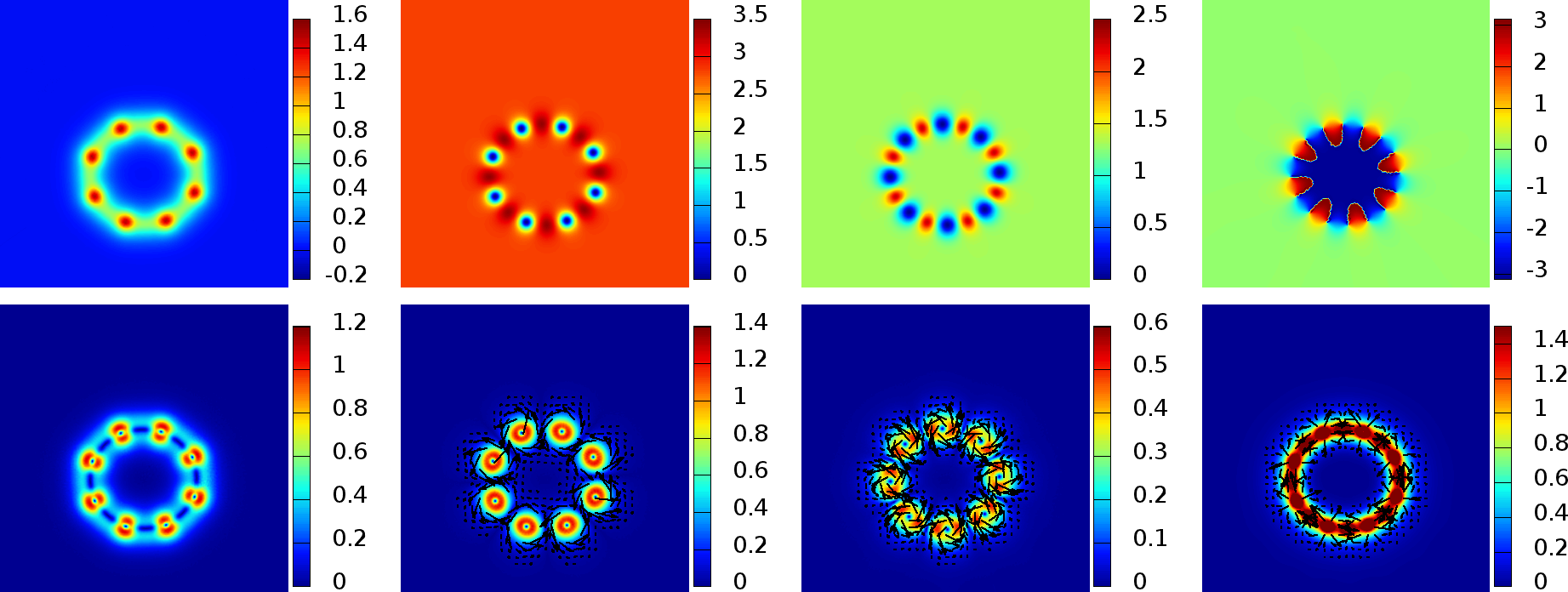}
 \hss}
\caption{
(Color online) -- 
$8$ vortex configuration. Parameters are $(\alpha_1,\beta_1)=(-3.6,1.0)$ 
and $(\alpha_2,\beta_2)=(-3.0,1.0)$ and $\gammaot=0.6$ with $e=0.6$. 
There is no Andreev--Bashkin coupling $\nuot=0.0$ but fractional 
vortices are split by bi-quadratic density coupling only. 
Displayed quantities are respectively the magnetic field $\B$, 
$|\psi_1|^2$, $|\psi_2|^2$ and the phase difference 
$\varphiot\equiv\varphi_2-\varphi_1$, on the first line. On the second 
line, $\J$, $\J_1$, $\J_2$ and ${\bf z}\times\Grad\varphiot$.
}
\label{Fig:Circle1}
\end{figure*}
\begin{figure*}[!hb]
 \hbox to \linewidth{ \hss
 \includegraphics[width=.8\linewidth]{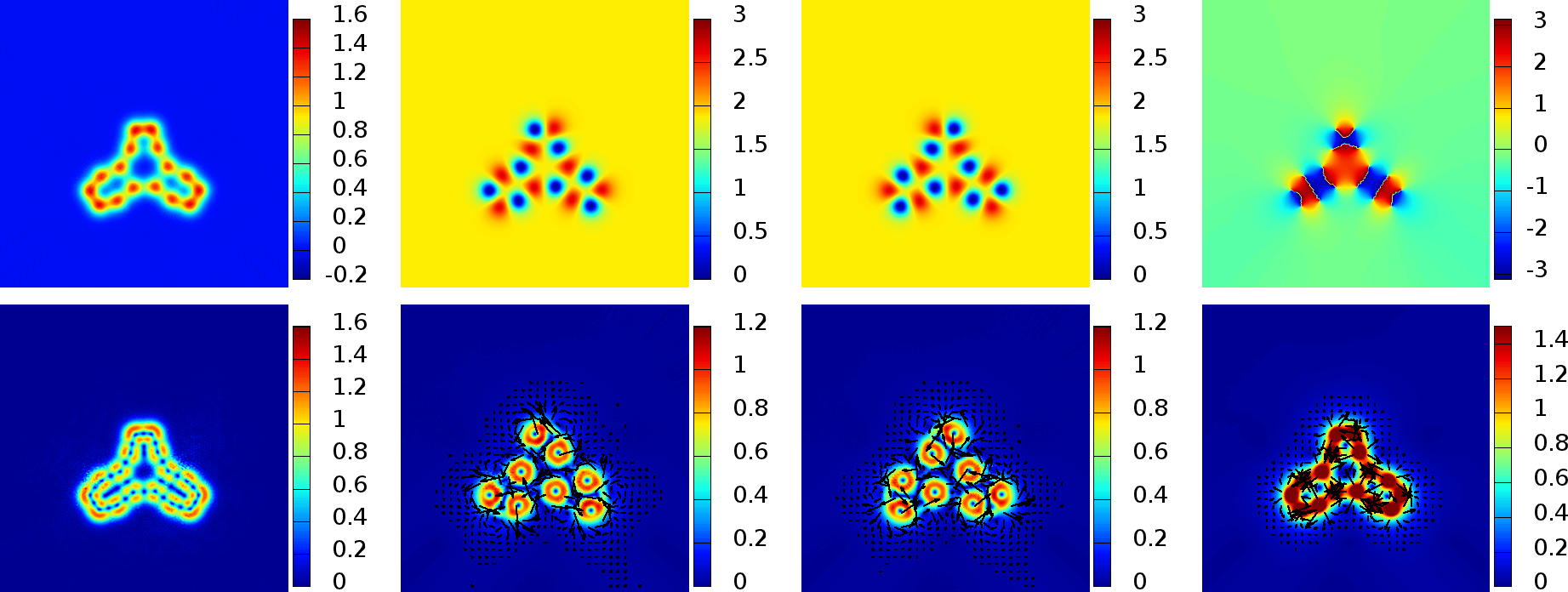}
 \hss}
\caption{
(Color online) -- 
Multiskyrmion carrying $\Q=8$ flux quanta, for identical components 
$(\alpha_a,\beta_a)=(-3.0,1.0)$ and $\gammaot=0.6$ with $e=0.8$. 
The Andreev--Bashkin coupling is $\nuot=1.0$. 
Displayed quantities are the same as in \Figref{Fig:Circle1}
}
\label{Fig:3lines}
\end{figure*}

\section{Additional material}\label{AppAdditional}

The bi-quadratic density interaction \Eqref{AppPotential2}, in 
\Eqref{AppFreeEnergy} also induces core splitting of the fractional 
vortices, for positive couplings $\gammaot$. Unlike the drag term which 
induces splitting by energetically penalizing  co-flowing currents, 
bi-quadratic density coupling (with $\gammaot>0$) penalizes core overlap 
directly. Indeed, it is energetically preferable to have singularities 
in each component sitting in different positions. 
Such a term is in general possible in multicomponent systems. Note 
that when the coupling are strong, it is no more favourable to have 
coexisting condensates and the superfluid density of a given condensate 
is completely suppressed (\ie phase separation). 

\begin{figure*}[!htb]
 \hbox to \linewidth{ \hss
 \includegraphics[width=.8\linewidth]{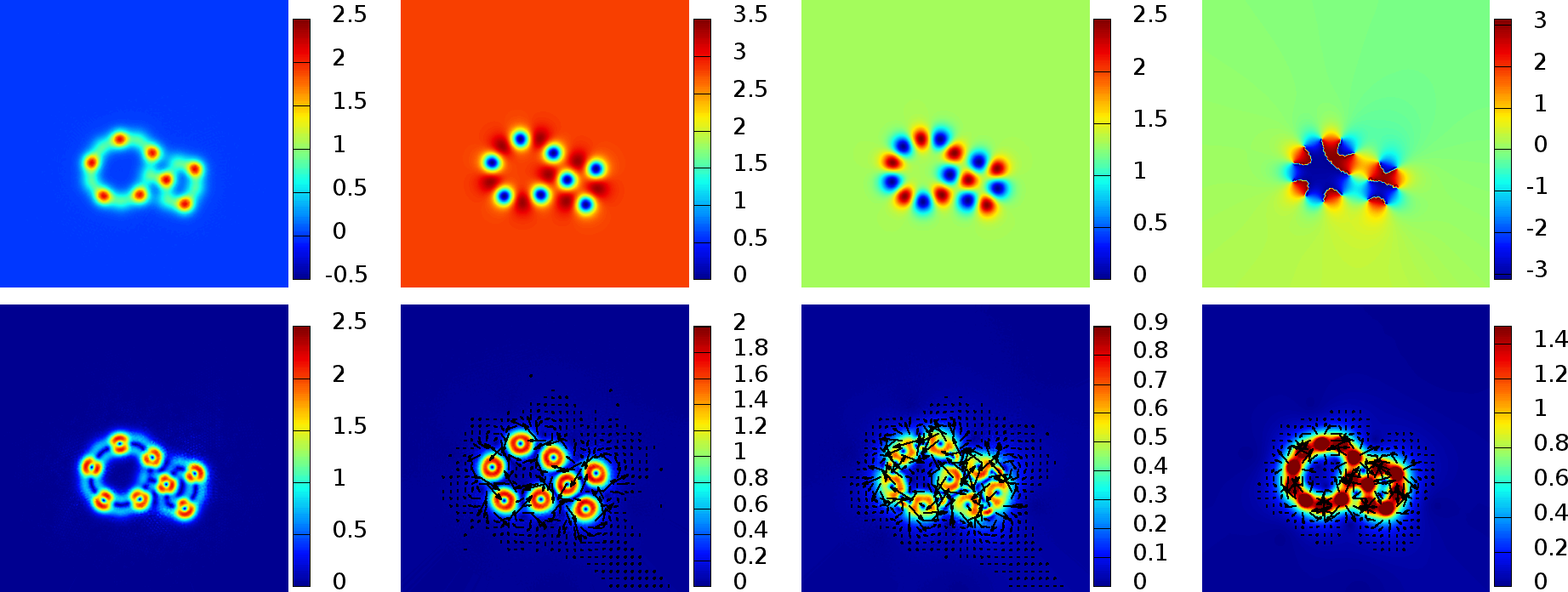}
 \hss}
\caption{
(Color online) -- 
A $8$ flux quanta configuration. Displayed quantities and the parameters 
are the same as in \Figref{Fig:Circle1} except for the coupling $\nuot=1$.
}
\label{Fig:Honeycomb2}
\end{figure*}

\begin{figure*}[!htb]
 \hbox to \linewidth{ \hss
 \includegraphics[width=.8\linewidth]{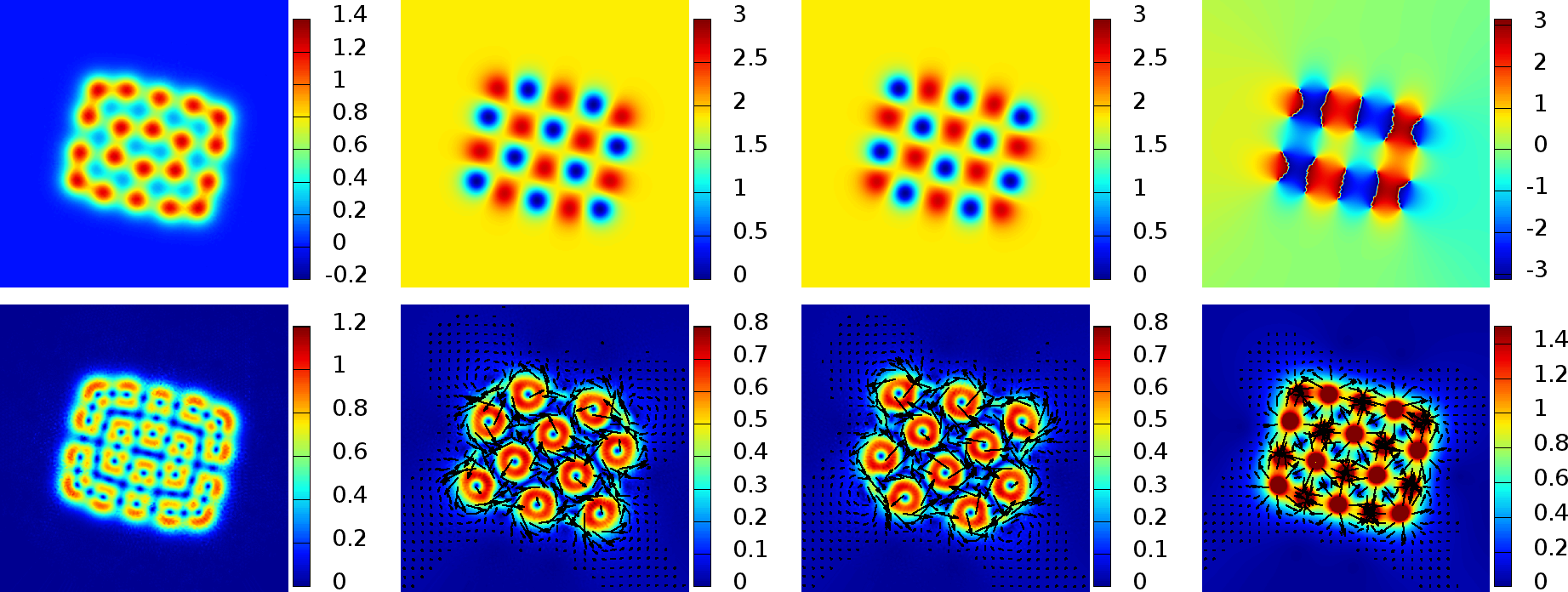}
 \hss}
\caption{
(Color online) -- 
A checkerboard cluster with $\Q=10$. Parameters are the same as in 
\Figref{Fig:3lines} except the gauge coupling $e=0.6$.
}
\label{Fig:Lattice1}
\end{figure*}

\begin{figure*}[!htb]
 \hbox to \linewidth{ \hss
 \includegraphics[width=.8\linewidth]{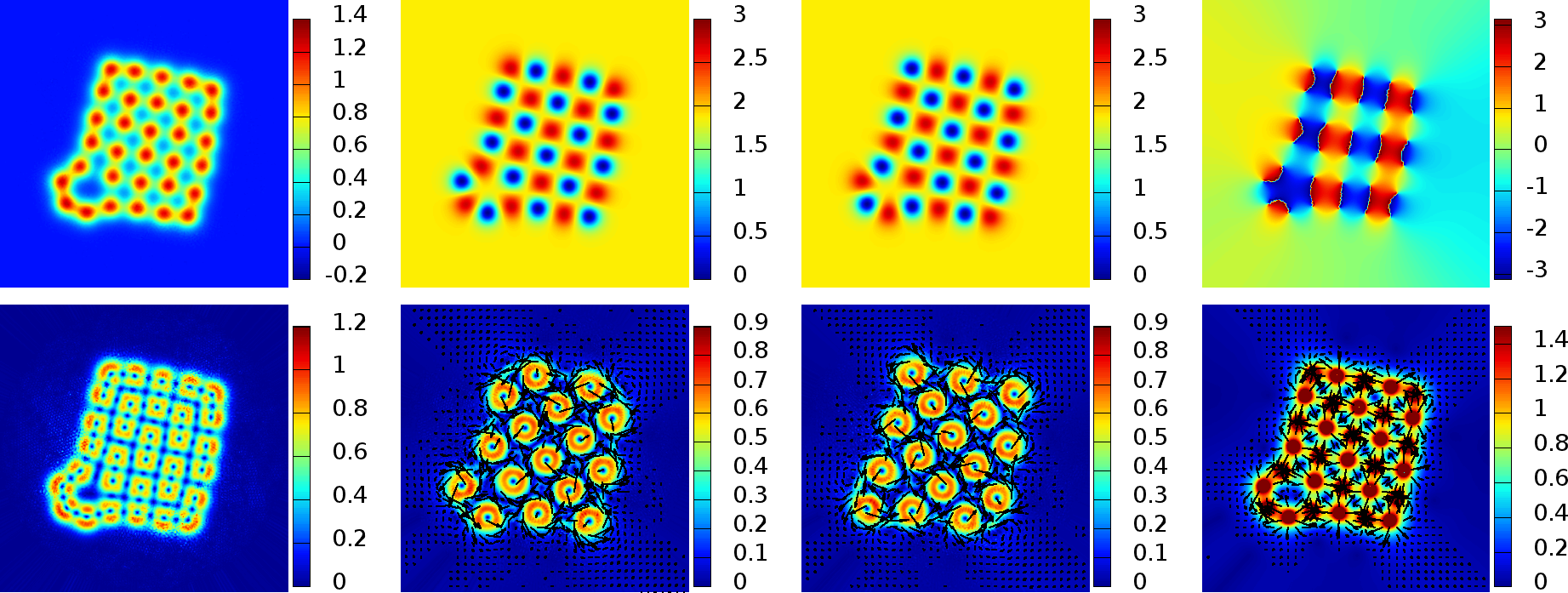}
 \hss}
\caption{
(Color online) -- 
A $\Q=16$ Skyrmions. The system compromises between the optimal 
compact packing and the number of vortices by creating a small loop 
at one of the corner. 
Parameters are the same as in \Figref{Fig:Lattice1}.
}
\label{Fig:Lattice2}
\end{figure*}

\begin{figure*}[!htb]
 \hbox to \linewidth{ \hss
 \includegraphics[width=.70\linewidth]{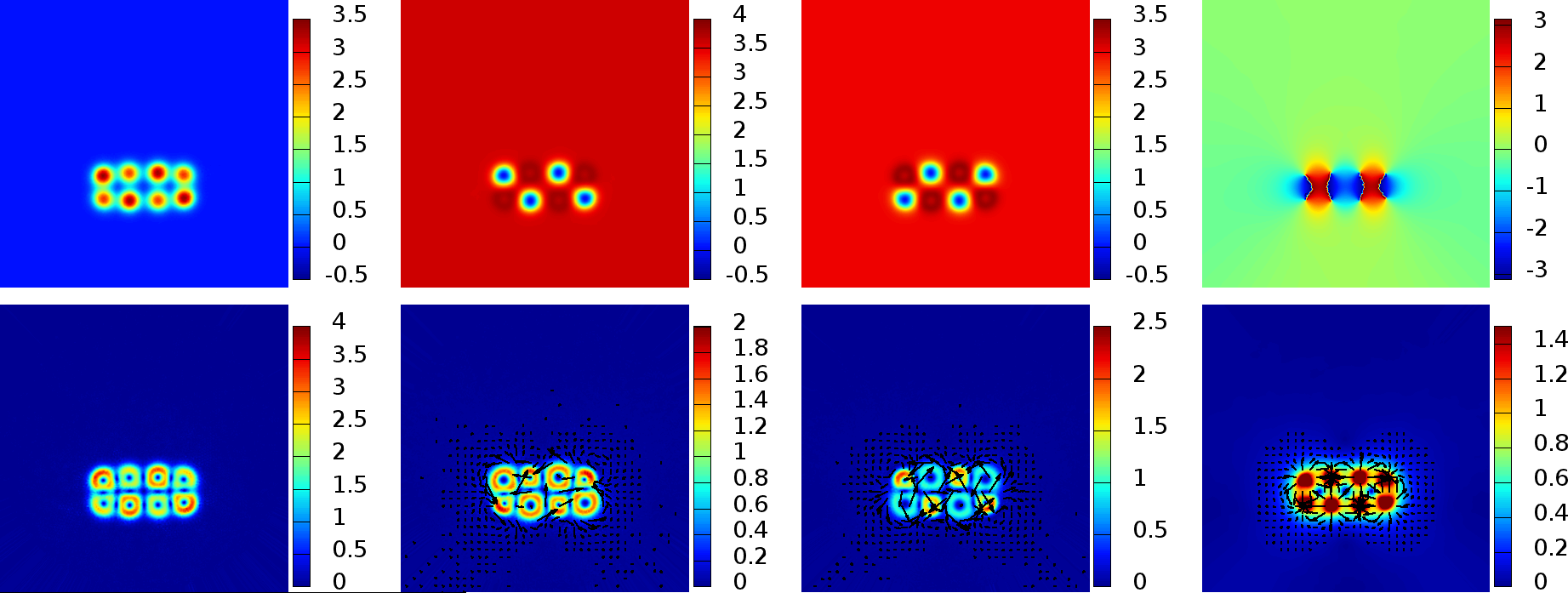}
\hspace{0.1cm}
 \includegraphics[width=.265\linewidth]{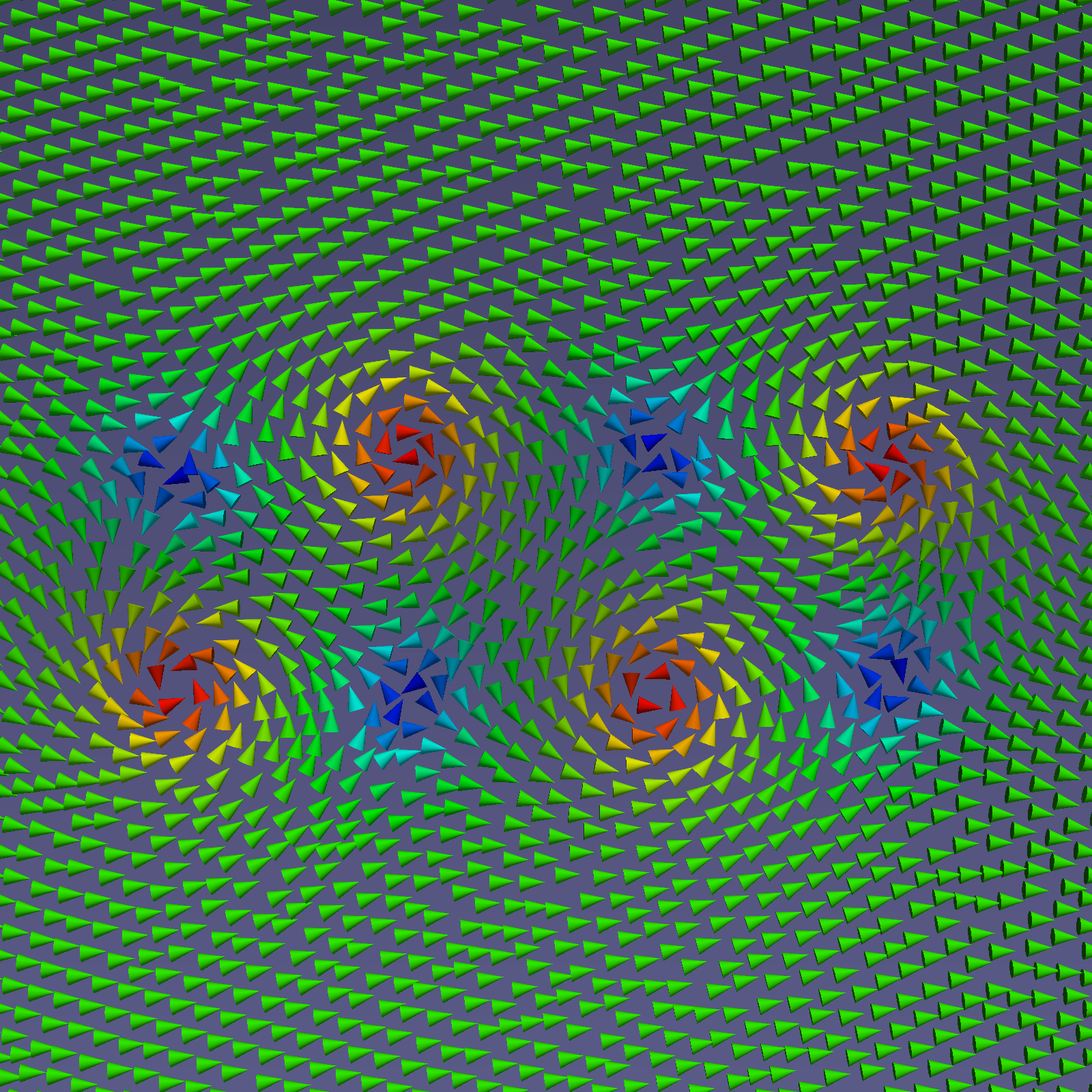}
 \hss}
\caption{
(Color online) -- A four quanta $\Q=4$ configuration. Parameters 
are $(\alpha_1,\beta_1)=(-3.6,1.0)$ $(\alpha_2,\beta_2)=(-3.0,1.0)$ 
with $e=0.3$ and Andreev--Bashkin coupling is $\nuot=5.0$. 
Bi-quadratic coupling vanish $\gammaot=0$.
Displayed quantities are the same as in \Figref{Fig:Honeycomb3}.
}
\label{Fig:Compact1}
\end{figure*}

Unlike the current drag interactions, the physics of the core splitting 
induced by bi-quadratic densities cannot be captured within the London 
limit (since it involves only densities). In  general  combining 
both dissipationless drag and bi-quadratic density interaction widely 
enriches the spectrum of various Skyrmionic structures which can be 
obtained. Figs.~\ref{Fig:Circle1}-\ref{Fig:Compact1} show detail of 
multiskyrmion solutions from the main body of the paper.

\section{Numerical Methods} \label{AppNumerics}

\subsection{Finite element energy minimization}

We consider the two-dimensional problem \Eqref{AppFreeEnergy} 
defined on the bounded domain $\Omega\subset\mathbbm{R}^2$ 
with $\partial\Omega$ its boundary. In practice we choose 
$\Omega$ to be a disk. The problem is supplemented by the 
boundary condition ${\bs n}\cdot\D\psia=0$ with ${\bs n}$ the 
normal vector to $\partial\Omega$. Physically this condition 
implies there is no current flowing through the boundary.
Since this boundary condition is gauge invariant, additional 
constraint can be chosen on the boundary to fix the gauge. 
Our choice is to impose the radial gauge on the boundary 
${\bs e}_\rho\cdot\A=0$ (note that with our choice of domain, 
this is equivalent to ${\bs n}\cdot\A=0$). With this choice, 
(most of) the gauge degrees of freedom are eliminated and 
the `no current flow' condition separates in two parts
\Equation{BC}{
   {\bs n}\cdot\Grad \psia=0~~~~~~\text{and}~~~~~~{\bs n}\cdot\A=0\,.
}
Note that these boundary conditions allow a topological defect 
to escape from the domain.
To prevent this in simulations of individual skyrmions or skyrmion
groups without applied field, the numerical grid is chosen to be 
large enough so that the attractive interaction with the boundaries 
is negligible for a given numerical accuracy. Thus in this method one 
has to use large numerical grids, which is computationally demanding. 
The advantage is that it is guaranteed that obtained solutions are 
not boundary pressure artifacts.

The variational problem is defined for numerical computation 
using a finite element formulation provided by the {\tt Freefem++} 
library \cite{Hecht:12}. Discretization within finite element 
formulation is done via a (homogeneous) triangulation over $\Omega$, 
based on Delaunay-Voronoi algorithm. Functions are decomposed 
on a continuous piecewise quadratic basis on each triangle. 
The accuracy of such method is controlled through the number of 
triangles, (we typically used $3\sim6\times10^4$), the order of 
expansion of the basis on each triangle (2nd order polynomial basis 
on each triangle), and also the order of the quadrature formula for 
the integral on the triangles. 

Once the problem is mathematically well defined, a numerical 
optimization algorithm is used to solve the variational nonlinear 
problem (i.e. to find the minima of $\F$). We used here a nonlinear 
conjugate gradient method. The algorithm is iterated until relative 
variation of the norm of the gradient of the functional  $\F$ with 
respect to all degrees of freedom is less than $10^{-6}$. 

\subsubsection*{Initial guess}

The initial field configuration carrying $N$ flux quanta is prepared 
by using an ansatz which imposes phase windings around spatially 
separated $N$ vortex cores in each condensates. 
\Align{Initial_Guess1}{
\psi_a&= |\psi_a|\mathrm{e}^{ i\Theta_a} \, ,~~  \nonumber \\
|\psi_a| &= u_a\prod_{k=1}^{N_v} 
\sqrt{\frac{1}{2} \left( 1+\tanh\left(\frac{4}{\xi_a}({\cal R}^a_k(x,y)-\xi_a) 
\right)\right)}\, ,
}
where $a=1,2\,$ and $u_a\,$ is the ground state value of each 
condensate density. The parameters $\xi_a$ parametrize the 
core size while 
\Align{Initial_Guess2}{
\Theta_a(x,y)&=\sum_{k=1}^{N}
      \tan^{-1}\left(\frac{y-y^a_k}{x-x^a_k}\right)  \,,\nonumber\\
{\cal R}^a_k(x,y)&=\sqrt{(x-x^a_k)^2+(y-y^a_k)^2}\,. 
}
$(x^a_k,y^a_k)$ determines the position of the core of $k$-th 
vortex of the $a$-condensate. The starting configuration of the 
vector potential is determined by solving Amp\`ere's law equation 
on the background of the superconducting condensates specified 
by \Eqref{Initial_Guess1}--\Eqref{Initial_Guess2}. Being a linear 
equation in $\A$, this is an easy operation.

Once the initial configuration defined, all degrees of freedom are 
relaxed simultaneously, within the `no current flow' boundary 
conditions discussed previously, to obtain highly accurate solutions 
of the Ginzburg-Landau equations. 

\subsection{Finite difference simulations}

In our simulations using finite differences, the energy functional 
\Eqref{AppFreeEnergy} is discretized in a gauge-invariance preserving 
manner using forward differences. For details of the discretization 
scheme, see \cite{juhasearlier}.
The constant applied external magnetic field ${\bs H}=H\ez$, 
is fixed by taking advantage of Stokes's theorem and specifying 
that $\A$ on the boundary satisfies 
\Equation{FD-bc}{
  \Curl\A = {\bs H}\,.
}
Stokes's theorem then ensures the flux through the system is equal 
to $\int_{\Omega \subset \mathbbm{R}^2} {\bs H}\cdot {\bs dS}$, but 
allowing $\A$ and hence $\B$ to vary arbitrarily inside the system. 
Note that this leaves gauge degrees of freedom in the system. However, 
in an energy minimization problem the algorithm only considers the 
energy which is a gauge-invariant quantity. Thus the possibility of 
evolving simply by a gauge transformation is eliminated since it does 
not lower the energy. The boundary condition is the discrete equivalent 
of ${\bs n}\cdot\D\psi_a=0$ and ensures that no supercurrent escapes 
the sample. This boundary is located several lattice points inside the 
computational lattice. This is the boundary of the sample and outside it, 
$\psi_i$ are not solved for.

The lattice parameters, $h_i$, control the accuracy of the lattice 
approximation and the minimization algorithm is considered to be 
converged whenever the largest discrete gradient in the system is 
below $10^{-5}\Pi_i h_i$ or the sup-norm of the discrete gradients is 
below $10^{-7}$. Some control calculations with a more restrictive 
convergence criterion were made but with no appreciable change to the 
solutions.

We typically used domains of $401\times403$ lattices points with 
lattice spacing of $h_i=0.1$. As an initial configuration, we set 
$\psi_a=0$ outside the superconductor (these values are not part of 
the minimization process), $\A=0$ everywhere, and 
$\psi_a=\sqrt{\tfrac{\alpha_a}{\beta_a}} \exp{i \varphi_a(x,y)}$, 
where phases $\varphi_a(x,y) \in [-\pi,\pi)$ are randomly chosen. 
At the beginning, therefore, we have $\B=0$ and this corresponds 
to a zero-field-cooled sample. When we have found a solution at a 
given external field, the boundary condition for $\A$ is updated 
to reflect the new field and the old solution is used as an initial 
guess for the next solution. A quasi-Newton algorithm with BFGS 
Hessian updates is used to simultaneously solve for all degrees of 
freedom subject to the boundary conditions at the two different 
boundaries (one for $\A$ and one for $\Psi$). The program itself is 
an extension of the one used in \cite{Palonen.Jaykka.ea:12} 
(for further details, see \cite{Palonen.Jaykka.ea:12} and the 
relevant references therein).


\subsection{Monte-Carlo simulations}

In the Monte Carlo simulations, vortices are treated as a system 
of $N$ point particles of two different colors, interacting with 
potentials \Eqref{AppInteraction}. The point particles live in a 
two-dimensional box $L\times L$ so that the number of particles per 
surface area is $N/L^2$. Periodic boundary conditions are imposed and 
the interaction is cut at half the box width. Tests with open boundary 
conditions without a cut-off have been performed and no structural 
differences are noted as compared to low-density simulations with 
periodic boundary conditions. Data are acquired during at least 
$10^4$ sweeps (a sweep constitutes a number of trial moves equal 
to the number of particles in the box), after an equilibration from 
a random initial configuration. The Monte Carlo trial moves consists 
of a single particle displacement, a pairwise displacement of a 
nearest-neighbours bound pair, or rotation of such a pair. The number 
of particles remains unchanged during the simulation. Furthermore, 
the maximal step length of a displacement is controlled such that 
approximately $10\%$ of the displacement trial moves are accepted. 
Parallel tempering is used in order for the low-temperature simulations 
to quickly converge into ordered low-energy states, as a low temperature 
simulation of these systems can easily be trapped in a metastable state.

The square lattice order parameter is defined as
\Equation{Psi4}
{
   \Psi_4=\frac{1}{4N}\left\vert\sum_{i=1}^N
\sum_{j=1}^4 \exp\left(4i\phi_{ij} \right)\right\vert \,,
}
where the sum in $j$ runs over the four nearest neighbors of particle $i$, 
and $\phi_{ij}$ is the angle of the line joining particles $i,j$ with
some arbitrary axis.

\input{Vortex-dipoles-v5.bbl}

\end{document}

%% file: Vortex-dipoles-v5.bbl
%